\documentclass[preprint]{sig-alternate-05-2015}
\usepackage{amsmath}
\usepackage{hhline}
\usepackage{url}
\usepackage{adjustbox}
\usepackage{multirow}
\usepackage{graphicx}
\usepackage{balance}
\usepackage[normalem]{ulem}
\usepackage{paralist}
\usepackage[backend=bibtex,firstinits=true,style=numeric-comp,sortcites=true,doi=false,isbn=false,url=false]{biblatex}

\newcommand{\RQone}{When does \textit{uniform} random mutant selection achieve an acceptable degree of representativeness of the full set of mutants?} %
\newcommand{\RQtwo}{To what extent can we reduce the acceptable sampling rate while keeping the same degree of representativeness?}
\newcommand{\RQthree}{How does the level of test adequacy  affect the acceptable sampling rate?} 

\begin{document}

\title{Evaluating Random Mutant Selection at Class-Level in Projects with Non-Adequate Test Suites}

\numberofauthors{1} %
\author{
	\alignauthor
	Ali Parsai, Alessandro Murgia, and Serge Demeyer\\
	\affaddr{University of Antwerp}\\
		\affaddr{Middelheimlaan 1}\\
		\affaddr{2020 Antwerp, Belgium} \\
		\email{\{ali.parsai, alessandro.murgia, serge.demeyer\}@uantwerpen.be}
		}

\date{19 January 2016}

\toappear{This is the preprint version of the article. Published final version can be found at \url{http://dl.acm.org/citation.cfm?id=2915992}}

\maketitle

\thispagestyle{plain}
\pagestyle{plain}

\begin{abstract}
Mutation testing is a standard technique to evaluate the quality of a test suite. Due to its computationally intensive nature, many approaches have been proposed 
to make this technique feasible in real case scenarios. Among these approaches, uniform random mutant selection has been demonstrated to be simple and promising. 
However, works on this area analyze mutant samples at project level mainly on projects with adequate test suites.
In this paper, we fill this lack of empirical validation by analyzing  random mutant selection at class level %
on projects with non-adequate test suites. 
First, we show that uniform random mutant selection underachieves the expected results. Then, we propose a new approach named weighted random mutant selection which generates more representative mutant samples. Finally, we
show that 
 representative mutant samples are larger for projects with high test adequacy.

\end{abstract}

\section{Introduction}

The quality of a test suite is of interest to researchers and practitioners since the early days of software testing. One of the extensively studied approaches to quantify the quality of a test suite is mutation testing~\cite{DeMillo1978}. %
Mutation testing provides a repeatable and scientific approach to measure the quality of the test suite,  
and it is proven to simulate the faults realistically~\cite{Andrews2005,Just2014}. This is due to the fact that the faults introduced by each mutant are modeled after the common mistakes developers often make~\cite{Jia2011}. 

Although the idea of mutation testing has been introduced since 1977~\cite{Hamlet1977,DeMillo1978}, it has not found widespread use in real scenarios due to its computationally intensive nature. 
Therefore several approaches have been proposed in order to make this technique feasible in industrial settings~\cite{Offutt2001}.
Among these approaches, random mutant selection is one of the easiest to implement with promising results~\cite{Wong1995}.
In this approach, instead of using all of the generated mutants, only a randomly selected subset of all mutants is selected to perform the mutation testing. 

A common way to compare the random mutant selection approaches is to evaluate their level of \textit{effectiveness}. The effectiveness of the sampled set is generally calculated in two steps. First, a certain number of test suites are created, each capable of killing all mutants in the sampled set. Then, the mutation coverage of each test suite is calculated, and the effectiveness of the sampled set is evaluated as the average of the mutation coverage of all these test suites~\cite{Zhang2010}. Using this procedure, the random mutant selection has been demonstrated to be effective in literature~\cite{Wong1993,Wong1995, Zhang2010,Zhang2013}. Here the mutant selection is performed using a uniform distribution, that is to say, all mutants had the same chance of being selected in the sampled set. Wong et al. reported that even using a subset of 10\% of all generated mutants there is only a decrease of 16\% of the performance achievable using full set of mutants~\cite{Wong1993,Wong1995}. Likewise, Zhang et al. reported that a sample reduced to half the size of the full set of mutants is equally effective~\cite{Zhang2010}. However, these studies suffer from two shortcomings. First of all, effectiveness gives a biased picture of the sampled set as it favors large classes with many mutants. Secondly, they presume that the test suite one starts from has a good mutation coverage. We explain these shortcomings in the next two paragraphs.

The effectiveness measure gives a biased view, because it is computed at project level and provides no information on how the mutation coverage of individual classes is affected by using the sampled set.  There might be classes where no mutants are selected at all, and yet the effectiveness of the sampled set would not be affected at project level. This is contrary to common practice, where test suite quality metrics are often calculated per class~\cite{Yang2009}. In that sense, studies based on the metric effectiveness lack of an analysis of the \textit{representativeness} of the sampled set of mutants; namely, how using a sampled set of mutants influences the mutation coverage calculated for each class.

The assumption that projects have an adequate test suite (thus a test suite that has a 100\% mutation coverage), is not realistic in many real projects. In that sense, studies based on this assumption need to be replicated in realistic scenarios to verify their validity.  Only Zhang et al.~\cite{Zhang2013} address the second threat by performing an experiment using non-adequate test suites as well as adequate test suites. However, they only considered the overall mutation coverage in their criteria and did not investigate the effects of random mutant selection on the mutation coverage at class level.
 
In this study, we attempt to fill the lack of empirical evaluation on random mutant selection on real projects 
by following the Goal-Question-Metric approach~\cite{Basili1993}. 
We set as \textit{object} the process of random mutant selection in software projects with non-adequate test suites.  
Our \textit{purpose} is to evaluate the representativeness of  random mutant selection at class and project level. 
We evaluate the representativeness for software projects with different level of test adequacy. 
Then, we propose a new approach named weighted random mutant selection  with the purpose of improving %
 the representativeness  of the sampled set  of mutants at class level. We also investigate how the level of test adequacy varies along with
acceptable sampling rate; namely the rate at which representativeness of the sampled set reaches a certain ``acceptable'' level (section \ref{subsect:cc}). %
The \textit{viewpoint} is that of software testers and testing researchers, both are interested in finding smaller yet representative sets of mutants able to work in real case scenarios. 
The \textit{environment} of this study consists of  12 open-source projects.
For this reason, we pursue the following research questions:

\begin{itemize}
	\item \textbf{RQ1:} %
	\RQone %
	
	We  evaluate the representativeness of  uniform random mutant selection (from now on, referred to as the uniform approach) with various sampling  rates on 12  open-source projects. Even though we are in agreement with previous research on this topic~\cite{Wong1993,Wong1995,Zhang2010,Zhang2013} that using random mutant selection at low rates is effective in estimating project level mutation coverage, yet, the sampled set of mutants does not accurately represent the full set of mutants in estimating mutation coverage at class level.

		\item \textbf{RQ2:} %
		 \RQtwo
		 
We introduce a  simple heuristic to improve the representativeness of randomly selected mutants. We call the new approach \textit{weighted} random mutant selection (from now on, referred to as the weighted approach). This approach reduces the size of the sampled set at which we achieve acceptable mutant representativeness. 
	\item \textbf{RQ3:} \RQthree

We investigate the effects of test adequacy on the acceptable sampling rate. We discover that the more adequate the test suite is, 
the higher the acceptable sampling rate becomes. %

\end{itemize}

The rest of the article is structured as follows.
In Section~\ref{sec:background}, background information regarding the study is provided. 
In Section~\ref{sec:es}, the details of the setup of the case study is discussed.
 In Section~\ref{sec:ar}, the results are analyzed. 
In Section~\ref{sec:threats} we discuss the threats that affect the results. 
In Section~\ref{sec:rw}, we report the state of literature on this topic. 
Finally, we present the  conclusion in Section~\ref{sec:conclusions}.

\section{Background}
\label{sec:background}

This section provides background information  about what mutant sampling is,  and the tool  we use for our study.

\subsection{Mutant Sampling}
To make mutation testing practical, it is important to reduce the time it needs to run. One way to achieve this is to reduce the number of mutants. A simple approach to mutant reduction is to randomly select a set of mutants. This idea was first proposed by Acree~\cite{Acree1980} and Budd~\cite{Budd1980} in their PhD theses. 
To perform random mutant selection,  %
we do not need any  extra information regarding the context of the mutants. This makes easier the implementation of the mutation testing tools. %
Because of this, and the simplicity of  random selection procedure, %
its performance overhead  is negligible as well. %

The random mutant selection can be performed uniformly, meaning that each mutant has the same chance of being selected.
Otherwise, the random mutant selection can be enhanced  %
by using heuristics based on  the source code. 

The percentage of mutants that are selected determines the \textit{sampling rate} for random mutant selection. Using a fixed sampling rate is common in literature~\cite{Wong1995,Zhang2010,Zhang2013}. It is also possible to determine the sampling rate  %
dynamically while performing mutation testing. A method resembling the latter was proposed by Sahinoglu and Spafford to randomly select the mutants until the sample size becomes statistically appropriate~\cite{Sahinoglu1990}. %
They concluded that their model achieves better results due to its self-adjusting nature~\cite{Jia2011}.

\subsection{LittleDarwin}
To perform our analysis, we used the LittleDarwin\footnote{\url{http://littledarwin.parsai.net/}} mutation testing tool previously used by Parsai et al.~\cite{Parsai2015,Parsai2015a}. This tool can perform %
mutation testing in complex (or simple) environments. %

LittleDarwin creates mutants by manipulating source code, and keeps the information about generated mutants and the results of the analysis on a local database, allowing  to analyze the results further by using subsets of the final result. This also allows for the manual filtering of \textit{equivalent} mutants\footnote{We did not filter the equivalent mutants (see Section~\ref{sec:threats}).}, namely, the mutants that keeps  the semantics of the program unchanged, and thus cannot be killed by any test~\cite{Grun2009}.

In its current version, LittleDarwin supports mutation testing of Java programs with in total 9 mutation operators. 
These mutation operators are an adaptation of the  %
minimal set introduced by Offutt et al.~\cite{Offutt1996}. 
The description of each mutation operator along with an example can be found in Table~\ref{mutationoperators}. 

\begin{center}
	\begin{table}
		\centering
		\adjustbox{max width=\linewidth}
		{\begin{tabular}{|l||l|c|c|}
				\hline \multirow{2}{*}{\textbf{Operator}} & \multirow{2}{*}{\textbf{Description}} & \multicolumn{2}{c|}{\textbf{Example}} \\
				\hhline{~~--} & & \textbf{Before} & \textbf{After} \\ 
				\hline
				\hline AOR-B & Replaces a binary arithmetic operator & $a + b$  & $a - b$ \\ 
				\hline AOR-S & Replaces a shortcut arithmetic operator & $++a$ & $--a$ \\ 
				\hline AOR-U & Replaces a unary arithmetic operator & $-a$ & $+a$ \\ 
				\hline LOR & Replaces a logical operator & $a\,\&\,b$ & $a\,|\,b$ \\ 
				\hline SOR & Replaces a shift operator & $a >> b$ & $a << b$ \\ 
				\hline ROR & Replaces a relational operator & $a >= b$ & $a < b$ \\ 
				\hline COR & Replaces a binary conditional operator & $a\:\&\&\:b$ & $a\,||\,b$ \\ 
				\hline COD & Removes a unary conditional operator & $!\,a$  & $a$ \\ 
				\hline SAOR & Replaces a shortcut assignment operator & $a\:*= b$ & $a\:/= b$ \\ 
				\hline 
			\end{tabular}}
			\caption{LittleDarwin mutation operators}
			\label{mutationoperators}
		\end{table}
	\end{center}

\section{Case Study Setup}
\label{sec:es}

In this section, we provide information about the dataset we use (Subsection~\ref{subsect:sc}) and the criteria adopted to evaluate approaches for random selection (Subsection~\ref{subsect:cc}). 
Finally, we report the details of the algorithms used for the analysis (Subsection~\ref{subsect:ed}).

\subsection{Dataset}%
\label{subsect:sc}
We selected 12 open source projects for our empirical study (Table~\ref{table:cases}). 
The selected projects differ in size of their production code, test code, number of commits, and team size  to provide a wide range of possible scenarios. Moreover, they also differ in adequacy of the test suite 
based on statement, branch, and mutation coverage (Table~\ref{table:cases}). All selected projects are written in Java, which is a widely used programming language in industry~\cite{Garousi2013}.

\begin{table}
\centering

		\adjustbox{max width=\linewidth}{
\begin{tabular}{|c|c|c|c|c|c|c|c|c|}
\hline \multirow{2}{*}{\textbf{Project}} & \multirow{2}{*}{\textbf{Ver.}} & \multicolumn{2}{c|}{\textbf{Size (LoC)}} & \multirow{2}{*}{\textbf{\#C}} & \multirow{2}{*}{\textbf{TS}} & \multirow{2}{*}{\textbf{SC}} & \multirow{2}{*}{\textbf{BC}} & \multirow{2}{*}{\textbf{MC}}\\
\hhline{~~--~~~} &  & \textbf{Prod.} & \textbf{Test} &  &  & & &  \\ 
\hline
\hline Apache Commons CLI & 1.3.1   & 2665 & 3768  & 816 & 15 & 96\% & 93\% & 94.2\%\\ 
\hline JSQLParser & 0.9.4 & 7342 & 5909  & 576 & 19 & 81\% & 73\% & 93.6\%\\ 
\hline jOpt Simple & 4.8 &  1982 & 6084 & 297 & 14 & 99\% & 97\% & 91.7\% \\ 
\hline Apache Commons Lang & 3.4   & 24289 & 41758 & 4398 & 30 & 94\% & 90\% & 90.7\% \\ 
\hline Joda Time & 2.8.1  & 28479 & 54645 & 1909 & 42 & 90\% & 81\% & 81.7\% \\ 
\hline Apache Commons Codec & 1.10   & 6485 & 10782 & 1461 & 10  & 96\% & 92\% & 81.6\% \\ 
\hline VRaptor & 3.5.5 & 14111 & 15496  &3417& 65 & 87\% & 81\% & 81.2\%\\ 
\hline JGraphT & 0.9.1 & 13822 & 8180  & 1150& 31 & 79\% & 73\% & 69.4\%\\ 
\hline AddThis Codec & 3.3.0 & 3675 & 1342  & 249& 4 & 69\% & 63\% & 64.7\%\\ 
\hline PITest & 1.1.7 & 17244 & 19005  & 1044 & 19 &  79\% & 73\%  & 62.9\%\\ 
\hline JTerminal  & 1.0.1 & 687 & 250  & 8 & 2 & 66\% & 56\% & 60.0\%\\ 
\hline JDepend & 2.9.1 & 2460 & 1053   & 18 & 2 & 59\% & 52\% & 59.0\%\\ 
\hline
 \multicolumn{9}{c}{} \\
 \multicolumn{1}{l}{Acronyms:} & \multicolumn{8}{r}{Version (Ver.), Line of code (LoC), Production code (Prod.),} \\
 	\multicolumn{9}{r}{Number of commits (\#C),  Team size (TS), Statement coverage (SC),}\\
 	\multicolumn{9}{r}{Branch coverage (BC), Mutation coverage (MC)}
\end{tabular}

}%
	\caption{Relevant statistics of the selected projects}
	\label{table:cases}
\end{table}

\subsection{Evaluation Criteria}

\label{subsect:cc}

The union of mutants generated by each class defines the full set of mutants for the project.
The mutation coverage acquired using random mutant selection only has practical purpose 
if it is representative of the mutation coverage  calculated using the full set of mutants.
 Here, for representative we mean the correlation between the mutation coverage for the sampled set, and the full set.

Previous studies investigating how to reduce the full set of mutants used the metric \textit{effectiveness}~\cite{Wong1993,Offutt1993}. 
To calculate effectiveness, first a set of test suites ($T$) is created, that each member %
 is a test suite capable of killing all non-equivalent %
mutants in the sampled set. The mutation coverage of each test suite is then calculated using the full set of mutants, and effectiveness is defined as the average of the mutation coverage for all test suites in $T$. This metric only takes into account the effectiveness of the sampled set  at project level. 

In this work, 
we propose a new metric called \textit{representativeness} to evaluate the correlation of the mutation coverage for the sampled set, and the full set of the mutants.
To compute the representativeness %
of the random sample, we use Pearson's $\rho$ and Kendall's $\tau_b$ correlation coefficients. We first 
partition the  %
 set of mutants according to set of classes, then we calculate the mutation coverage for each class using the equation in Figure~\ref{eq:mutationscore}. %
The same procedure is performed using the sampled set of mutants. The results are then used to calculate correlation coefficients. 

\begin{figure}[!b]
\centering{ 
$Mutation\ Coverage = \dfrac{Killed\ Mutants}{All\ Mutants}$
}
\caption{Mutation coverage equation}
\label{eq:mutationscore}
\end{figure}

We analyze whether the values of mutation coverage are linearly correlated (using Pearson's $\rho$~\cite{Gayen1951}), and whether the ranking order of results can be predicted by the mutation coverage calculated using sampled sets (using Kendall's $\tau_b$ \cite{Abdi2007,Agresti2010}). Each one of these correlation coefficients provide a different outlook on the results. Pearson's $\rho$ evaluates the linear correlation between the mutation coverage values of each class. The higher the correlation between the two sets (sampled and full), the higher is the representativeness of the sampled mutants. Kendall's $\tau_b$ shows if the sampled set can accurately predict the ranking order of the classes based on mutation coverage. This is desired, for example, if prioritizing the classes based on mutation coverage is important for the user.  This coefficient has been used previously by Zhang et al.~\cite{Zhang2010,Zhang2013}. 
By using this criteria, we aim to compare two approaches from the viewpoint of a developer who intends to discover which class is in need of more testing.

\begin{figure}[!t]
	\centering
	\includegraphics[width=0.85\linewidth]{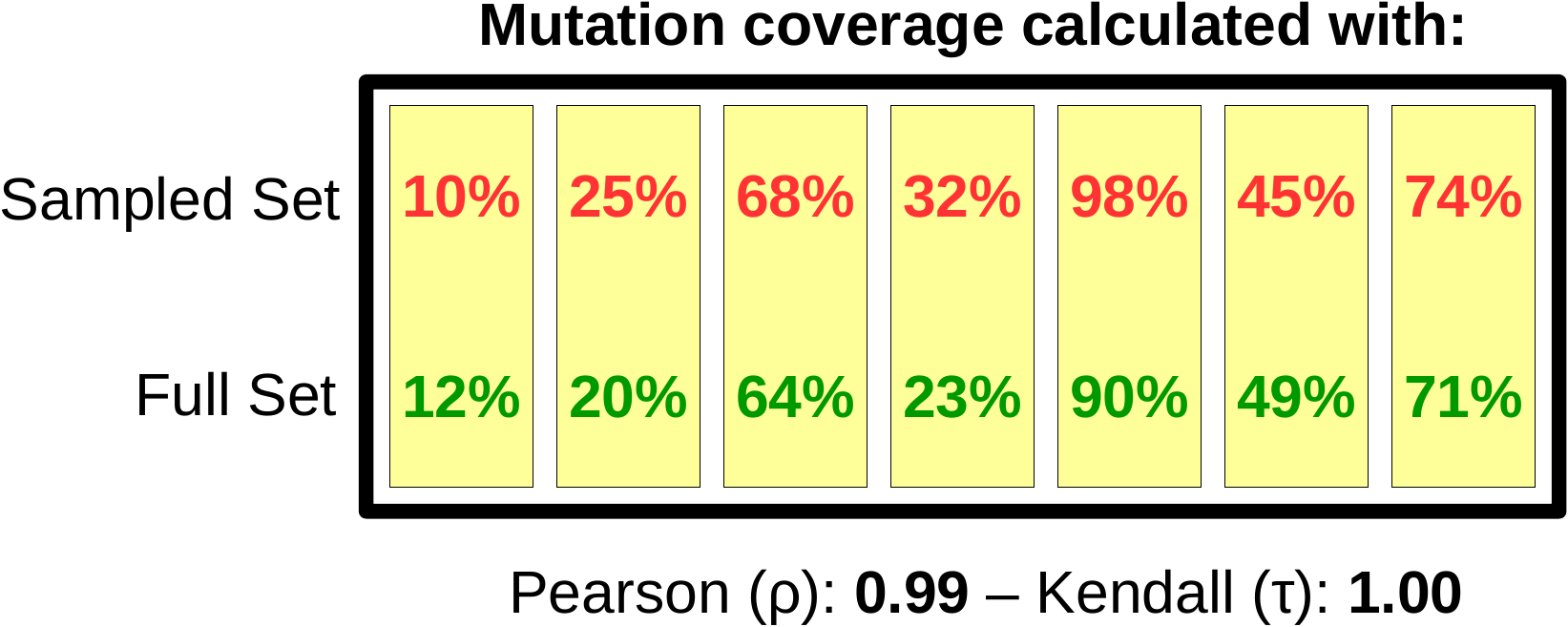}
	\caption{An example of calculation of correlation}
	\label{fig:criteriaexample}
\end{figure}

An example of these criteria is shown in Figure~\ref{fig:criteriaexample}. In this figure, the small rectangles represent the classes and the large rectangle represents the project. The percentages inside each rectangle shows the mutation coverage of that class. The mutation coverage calculated using sampled set is shown in red while the mutation coverage calculated using all mutants is shown in green. The correlation between these two sets of mutation coverage values are then calculated using $\rho$ and $\tau_b$.

Presenting mutation coverage as a percentage  follows the same trend as other metrics such as branch coverage~\cite{Zhu1997,Yang2009}. 
We use mutation coverage as a metric to evaluate the quality of the test suite in the same manner. 
For object oriented programming languages like Java, test suite quality metrics are often calculated per unit, meaning that tools calculate the coverage per smaller units than the project as a whole~\cite{Yang2009}. 
Therefore, it is important that the results acquired from sampled set can emulate the results of full set of mutants per class. %

We consider a correlation value ($\rho$)  of 0.75 the \textit{critical point} after which two sets of mutation coverage values are strongly correlated. The choice of critical point is only to provide a reference, and slightly higher or lower values will not affect our conclusions (See Section~\ref{sec:threats}). %
In our analysis, the size of these sets is always higher than 16 (granting a p-value lower than 0.01), with only the exception of JTerminal where the size is 6 (providing a p-value lower than 0.09).
From now on, we define \textit{acceptable} anything higher than the critical point, and as \textit{acceptable sampling rate} the acceptable rate from
where the degree of representativeness remains acceptable. %

In order to evaluate the representativeness of the sampled set at project level,

we calculate the difference between the mutation coverage from the full set and the mutation coverage from the sampled set at various sampling rates. 
From now on, we refer to this difference as ``distance'', namely the absolute value of the difference between mutation coverage of the full set and the mutation coverage of the sampled set.
This gives us an idea on how close the sampled set can approximate the overall mutation coverage at various rates.

\subsection{Algorithms}
\label{subsect:ed}
In order to calculate the representativeness of the sampled set of mutants, first we need to calculate the
 mutation coverage for each class using the sampled set. To do this, first we collect all the mutants belonging to the class from the sampled set. Then we use the equation in  Figure~\ref{eq:mutationscore} to calculate the mutation coverage for that class. This procedure is repeated until each class has two mutation coverage measurements: one calculated using the full set, and another calculated using the sampled set. Then we calculate the correlation between these two sets of mutation coverage values using the correlation coefficients $\rho$ and $\tau_b$. This process is repeated 10 times in order to reduce the random noise.
 Finally, the average of all correlation values is reported for a specific sampling rate. By varying the sampling rate from 1\% to 100\%, we find 
 the acceptable sampling rate for each project. %

 The process of selection of mutants is performed in two different ways for our analysis. For the uniform approach, we use a random function to select $N$ mutants from the full set.  
 In the weighted approach, we first assign a weight proportional to inverse of the size of the class to each mutant, and then we use a roulette wheel algorithm to select the mutants. 
Assigning the inverse of the size avoids the overrepresentation of larger classes over smaller ones. 
	
 To do this, we first pick a random number $r$ between 0 and  the sum of all weights. Then, we add the weight of each mutant until this sum is greater than the random number  $r$. The corresponding mutant is then selected. This procedure is repeated until the sample size reaches $N$. 
For the interested reader, the details of these algorithms are available online\footnote{\url{http://parsai.net/files/WRMSAFormalDef.pdf}}. %

\section{Results}
\label{sec:ar}
In this section, we discuss the results of our study. For each research question, we first briefly describe our motivation, approach, 
and then our findings. 
In Table~\ref{table:results}, for each project  we report 
the acceptable sampling rate as determined by $\rho$ and $\tau_b$.
We refer to Figures \ref{fig:pearson} and \ref{fig:kendall} to summarize the results of all research questions.
In these figures, the horizontal axis is the sampling rate, and the vertical axis is the degree of representativeness.
The red line shows the data for the uniform approach and the blue line shows the data for the weighted approach. 
We also draw a green line at the critical point to show where the sets of mutants start to be acceptable.
While performing the study, we realized that the results from Pearson's correlation and those of Kendall correlation were very close to each other. Therefore, it is unnecessary to report both in many parts of the analysis. So, wherever not specified, we refer to the Pearson's correlation to  report the degree of representativeness.

\subsection*{\textit{RQ1. \RQone}} 
 \textbf{Motivation}. 
For different sampling rates, we want to evaluate  the representativeness of the sampled set of mutants with respect to the full set of mutants.
 We want to perform this analysis at project and class level.

\textbf{Approach}. We perform an empirical study on 12 object oriented projects with different levels of test adequacy. For each project, 
we evaluate the degree of representativeness of the sampled set at project level. 
For this reason, we analyze  the distance between the mutation coverage calculated from the sampled set, and the full set for sampling 
rates between 1\% and 100\%. %
Then, we calculate the degree of representativeness of the sampled set of mutants for different sampling rates at class level. 

\begin{figure}
\centering
\includegraphics[width=\linewidth]{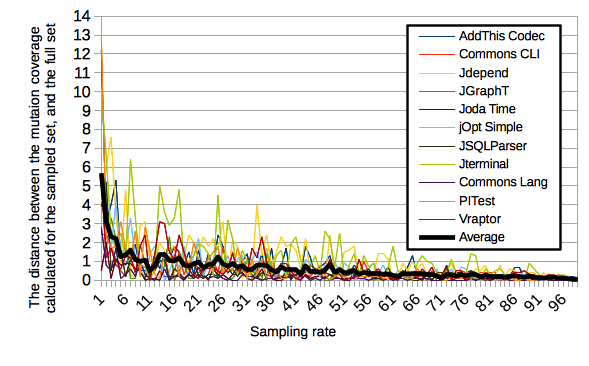}
\caption{The distance between mutation coverage calculated from sampled set and full set  for sampling rates from 1\% to 100\%}
\label{fig:overallcoverage}
\end{figure}

\textbf{Findings}. %
Zhang et al.~\cite{Zhang2013} states that the uniform  approach for adequate as well as  non-adequate test suites provides near-perfect results with a low sampling rate at project level. %
Figure~\ref{fig:overallcoverage} shows the distance %
between the mutation coverage calculated from sampled set and full set. In this figure, we observe that the average distance over all of our projects is below 2\% with a sampling rate as low as 5\%. 
Our analysis confirms Zhang's observation, namely \textbf{at project level,  the acceptable degree of representativeness is achieved at a very low  sampling rate.} %
The different level of representativeness achieved at project and class levels can be explained by the dominance of classes with larger sets of mutants in the overall mutation coverage. 
\textbf{At class level, the acceptable degree of representativeness is achieved for  sampling rates that span from 36\% to 88\%} (37\% to 83\% using Kendall's correlation).
Table \ref{table:results} shows that in JDepend, the appropriate acceptable sampling rate is 36\%, %
since from this rate we obtain an acceptable representativeness. %
For JSQLParser the acceptable sampling rate is 88\%.
These two projects represent the maximum variability of the acceptable sampling rate. %
On average, the acceptable sampling rate for the uniform approach is 63\%.

Focusing on Figures \ref{fig:pearson} and \ref{fig:kendall}, we can notice that for many projects the sampling rate has an almost linear relationship with the degree of representativeness (e.g.; VRaptor, PITest, and Commons Lang). Whereas, in some other projects, this relationship is logarithmic-like, with 
 the degree of representativeness that becomes acceptable sooner (e.g.; JGraphT, and JDepend). 
During our investigation, we found that correlation between the acceptable sampling rate and the number of classes with mutants is 0.44.
Visually, this correlation can be seen in Figure~\ref{fig:RQ3-1}. All these results show that \textbf{at class level, the acceptable sampling rate is project-dependent}.  Moreover, we can see that difference between the number of mutants in the sampled set  and the full set  does not justify the degree of representativeness. 

This result was unexpected since the analysis at project level reported that the uniform approach was viable with acceptable sampling rate as low as 5\%. Whereas, our analysis at
class level strongly differs having a acceptable sampling rate of  65\% on average. %
From this point of view,  the uniform approach underachieves the expected results.%

\subsection*{\textit{RQ2. \RQtwo}} 
\begin{figure*}
	\begin{minipage}{0.33\textwidth}
		\centering
		\includegraphics[width=\linewidth]{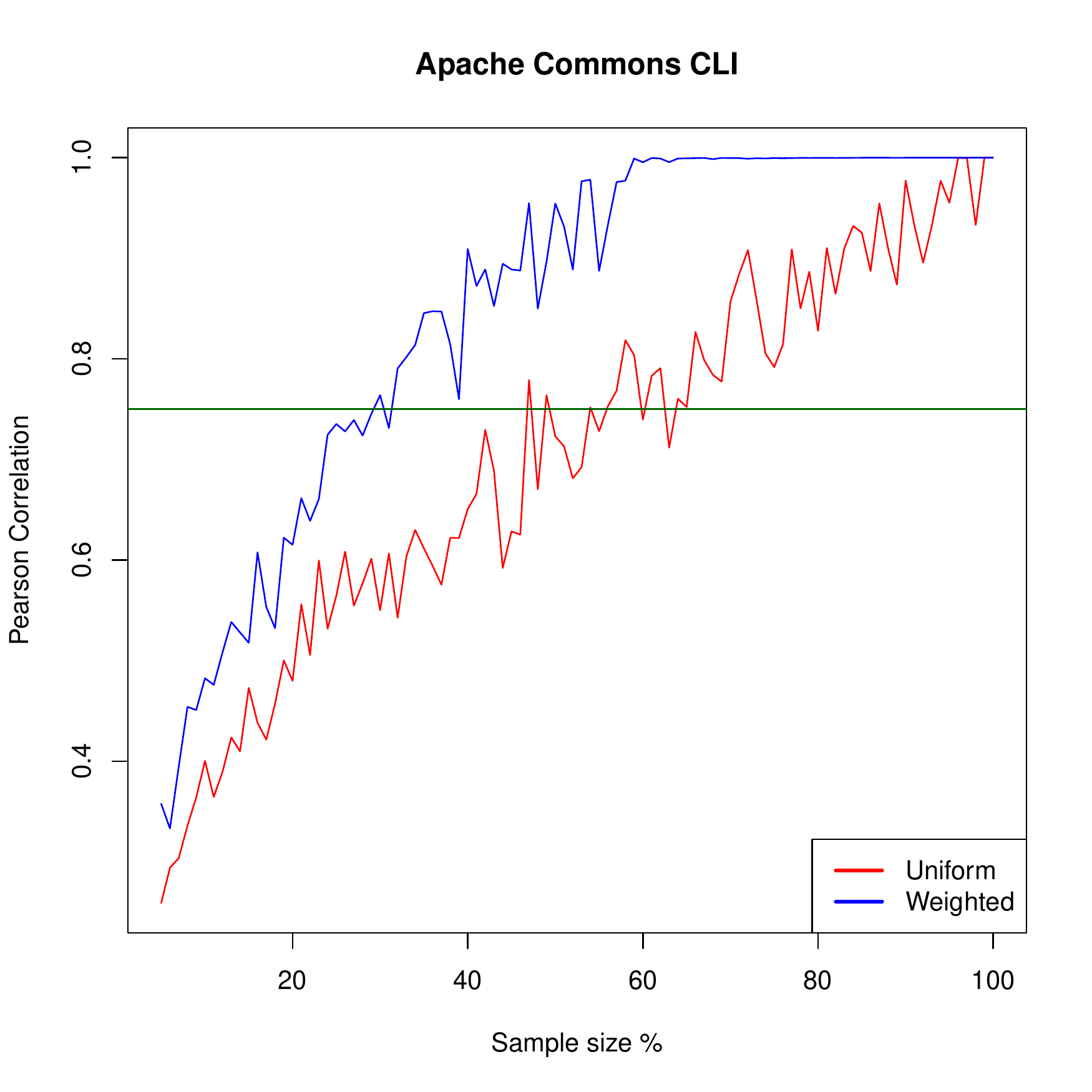}
	\end{minipage}
	\begin{minipage}{0.33\textwidth}
		\centering
		\includegraphics[width=\linewidth]{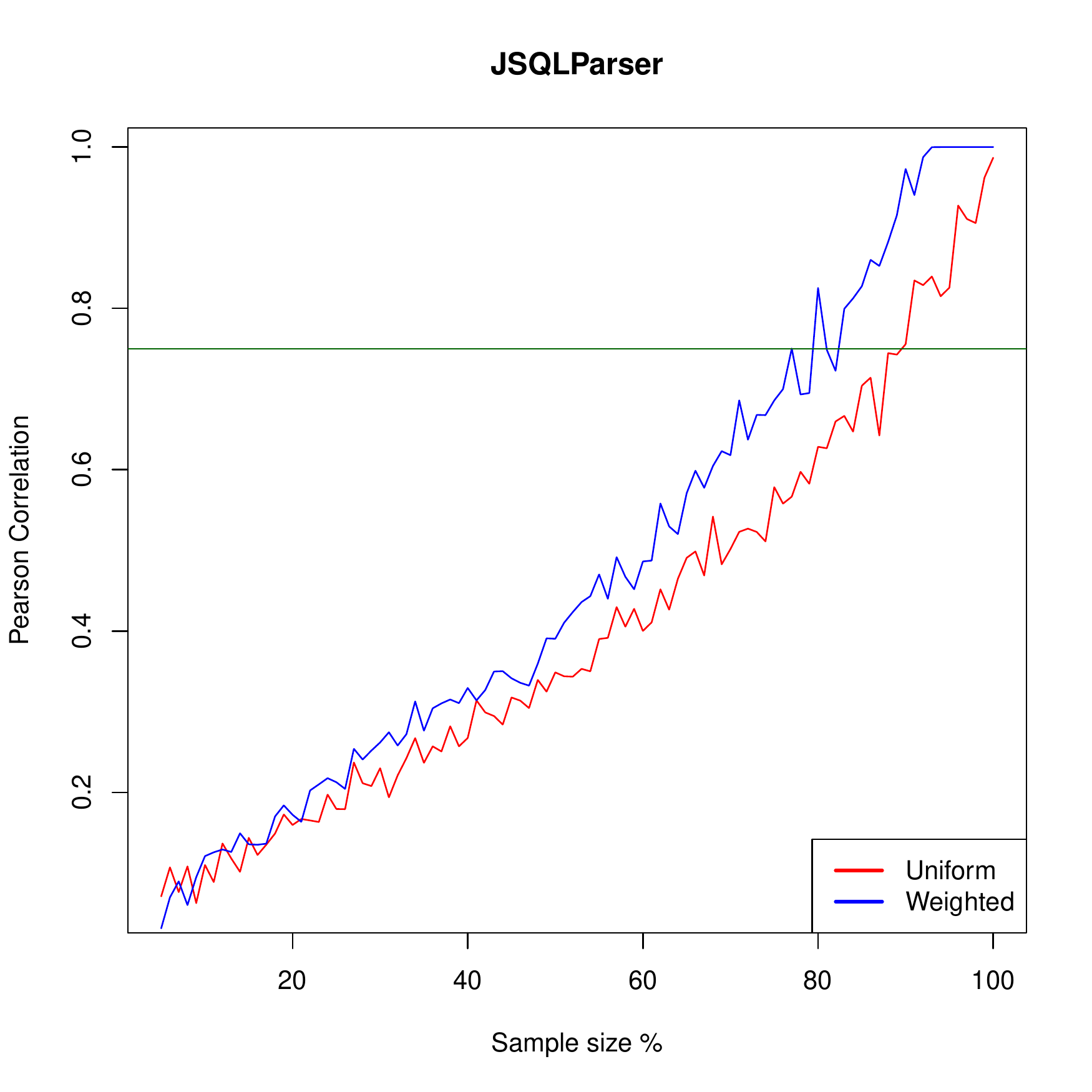}
	\end{minipage}
	\begin{minipage}{0.33\textwidth}
		\centering
		\includegraphics[width=\linewidth]{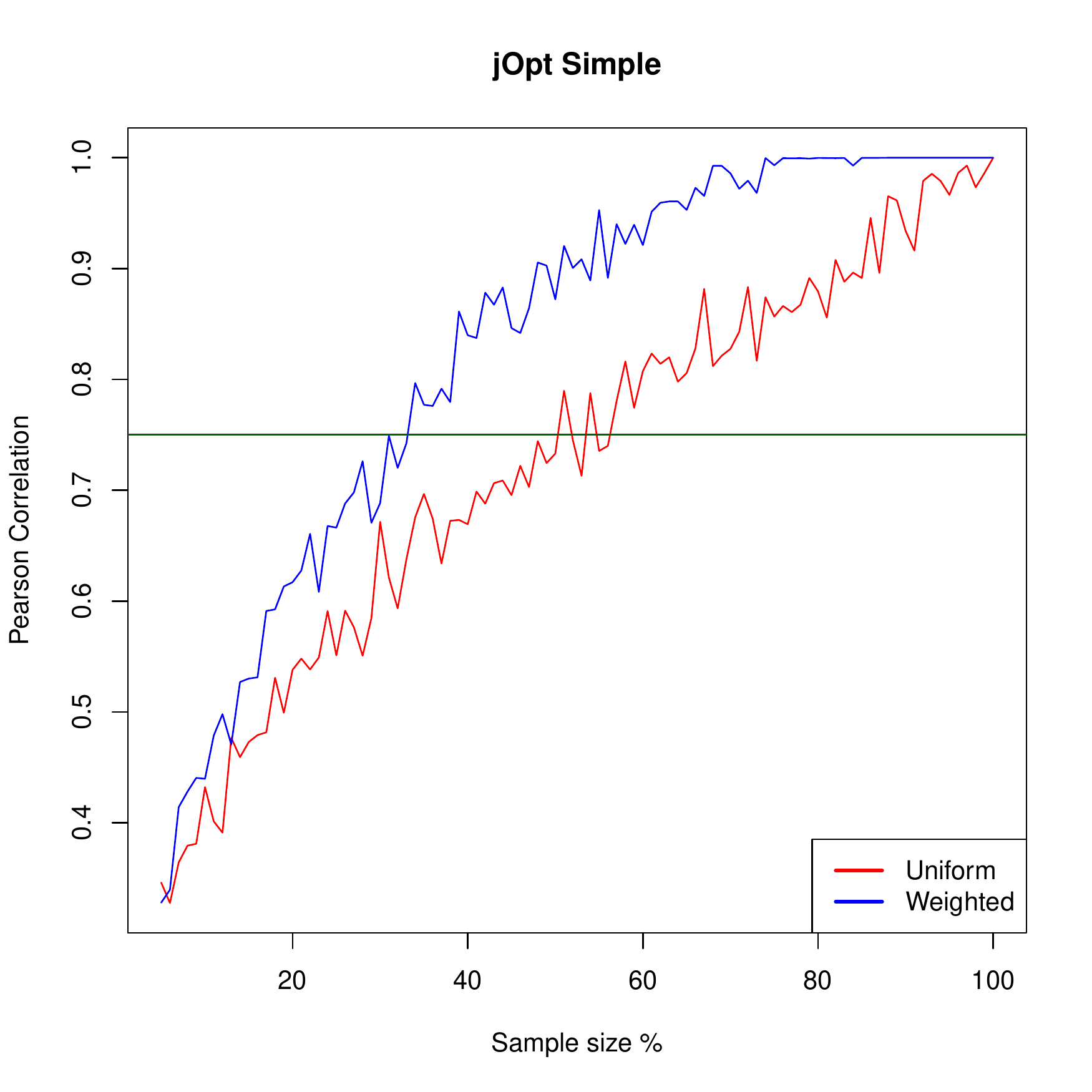}
	\end{minipage}
	\begin{minipage}{0.33\textwidth}
		\centering
		\includegraphics[width=\linewidth]{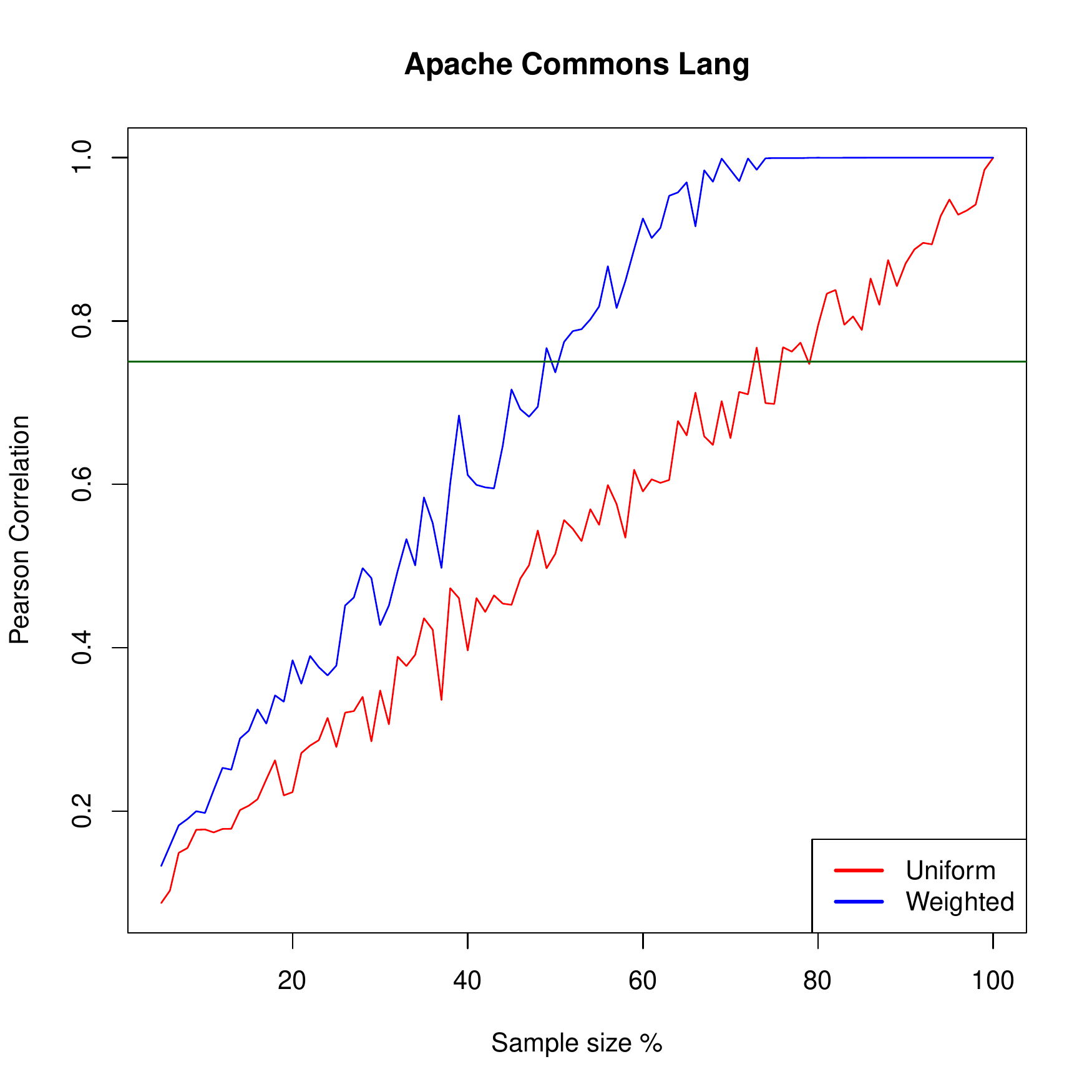}
	\end{minipage}
	\begin{minipage}{0.33\textwidth}
		\centering
		\includegraphics[width=\linewidth]{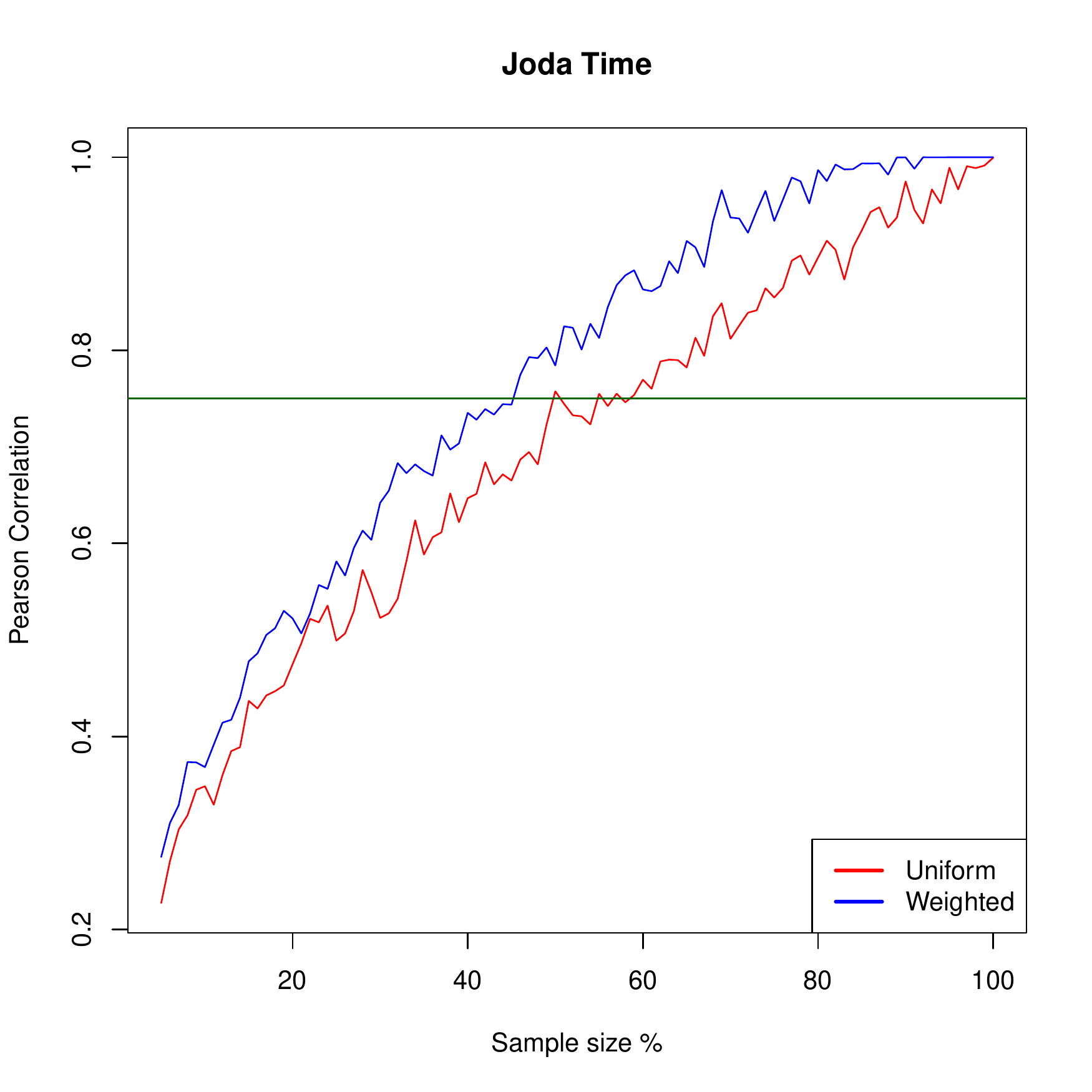}
	\end{minipage}
	\begin{minipage}{0.33\textwidth}
		\centering
		\includegraphics[width=\linewidth]{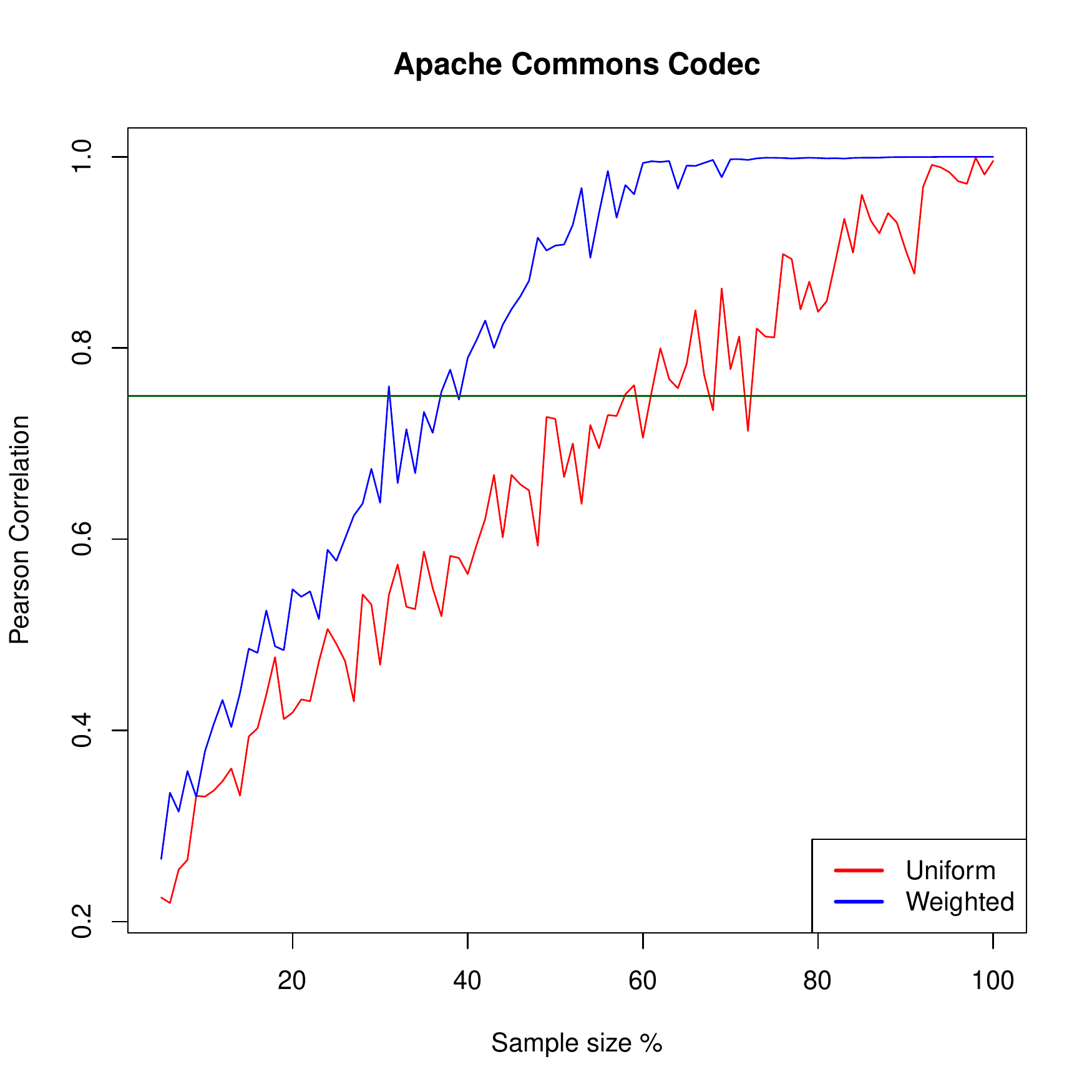}
	\end{minipage}
	\begin{minipage}{0.33\textwidth}
		\centering
		\includegraphics[width=\linewidth]{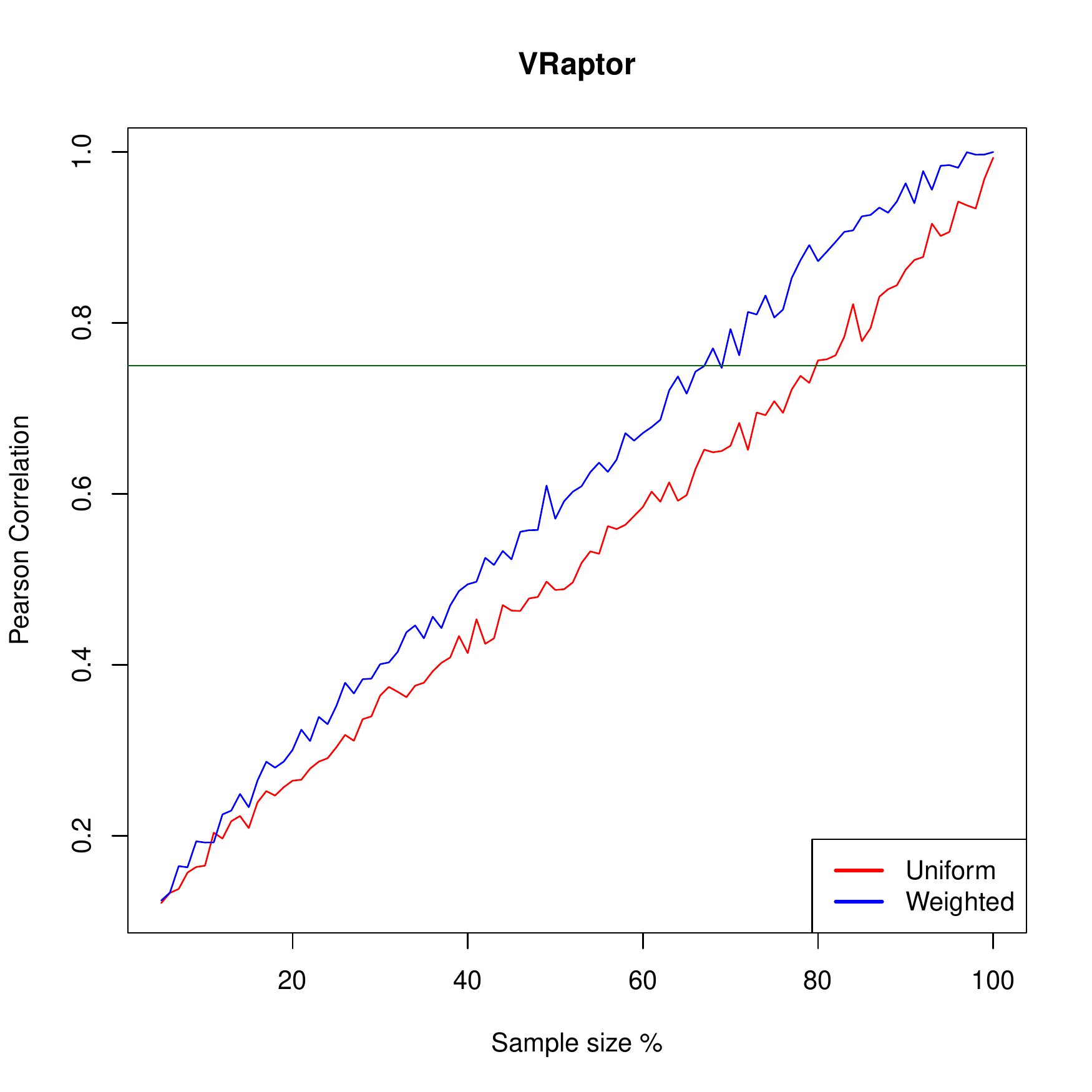}
	\end{minipage}
	\begin{minipage}{0.33\textwidth}
		\centering
		\includegraphics[width=\linewidth]{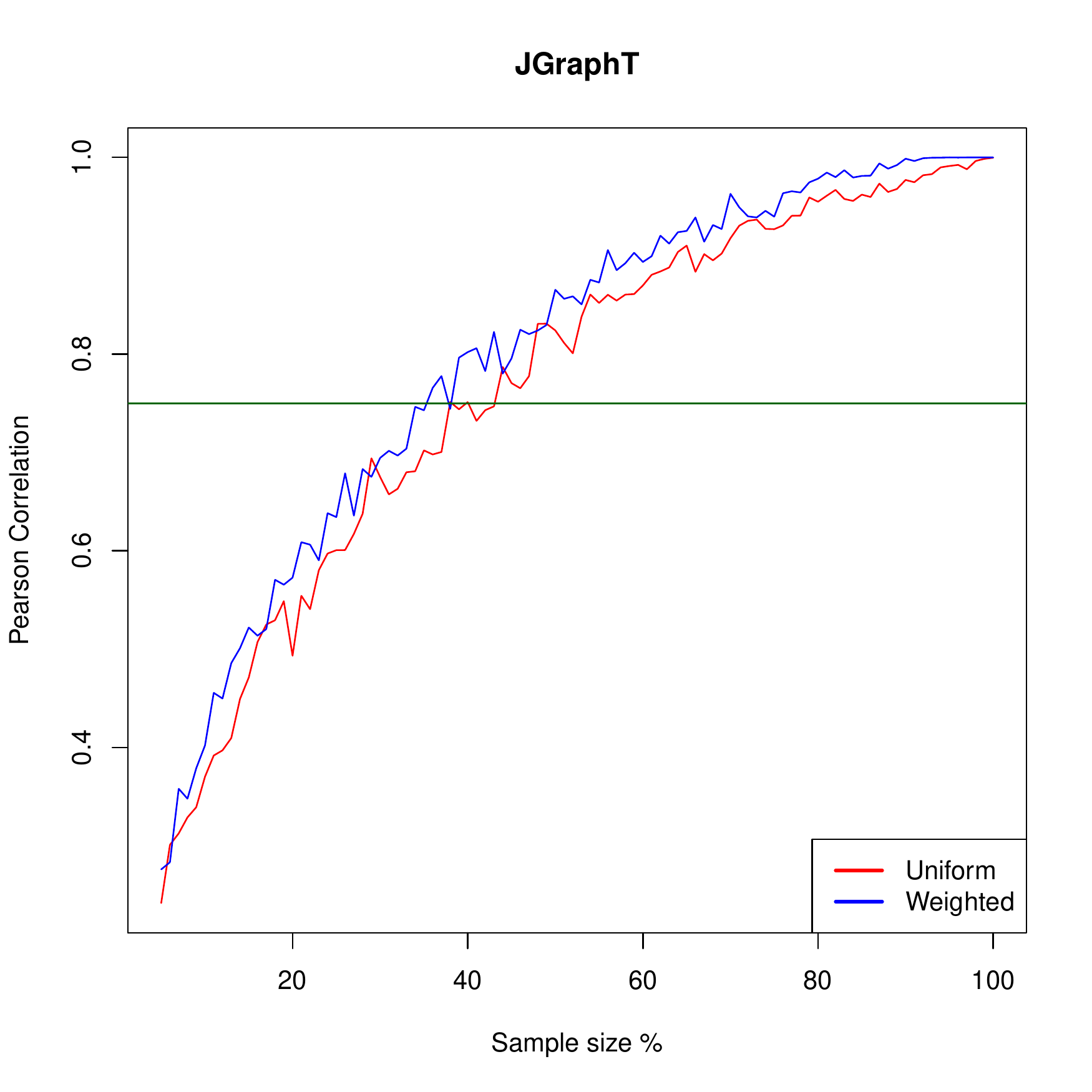}
	\end{minipage}
	\begin{minipage}{0.33\textwidth}
		\centering
		\includegraphics[width=\linewidth]{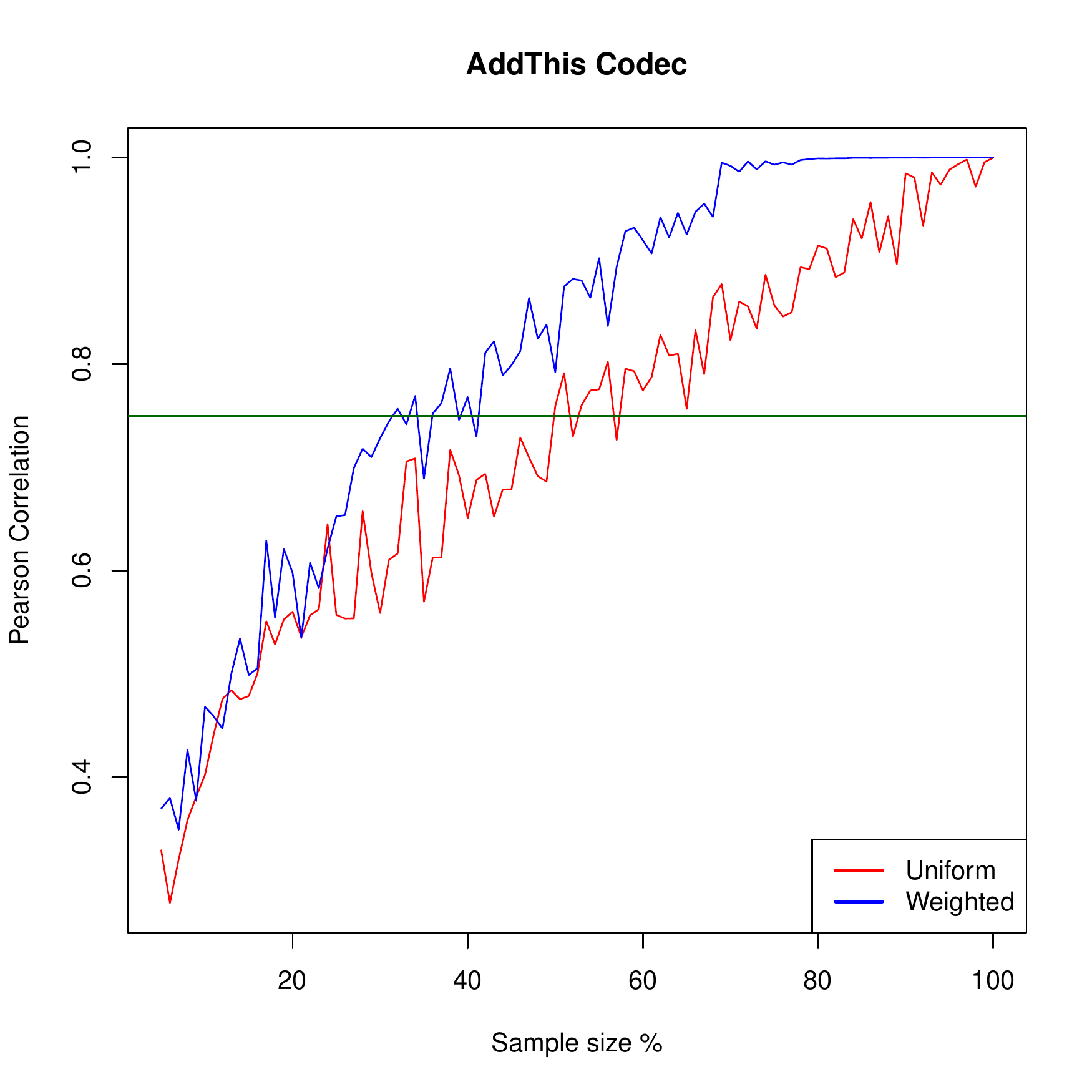}
	\end{minipage}
	\begin{minipage}{0.33\textwidth}
		\centering
		\includegraphics[width=\linewidth]{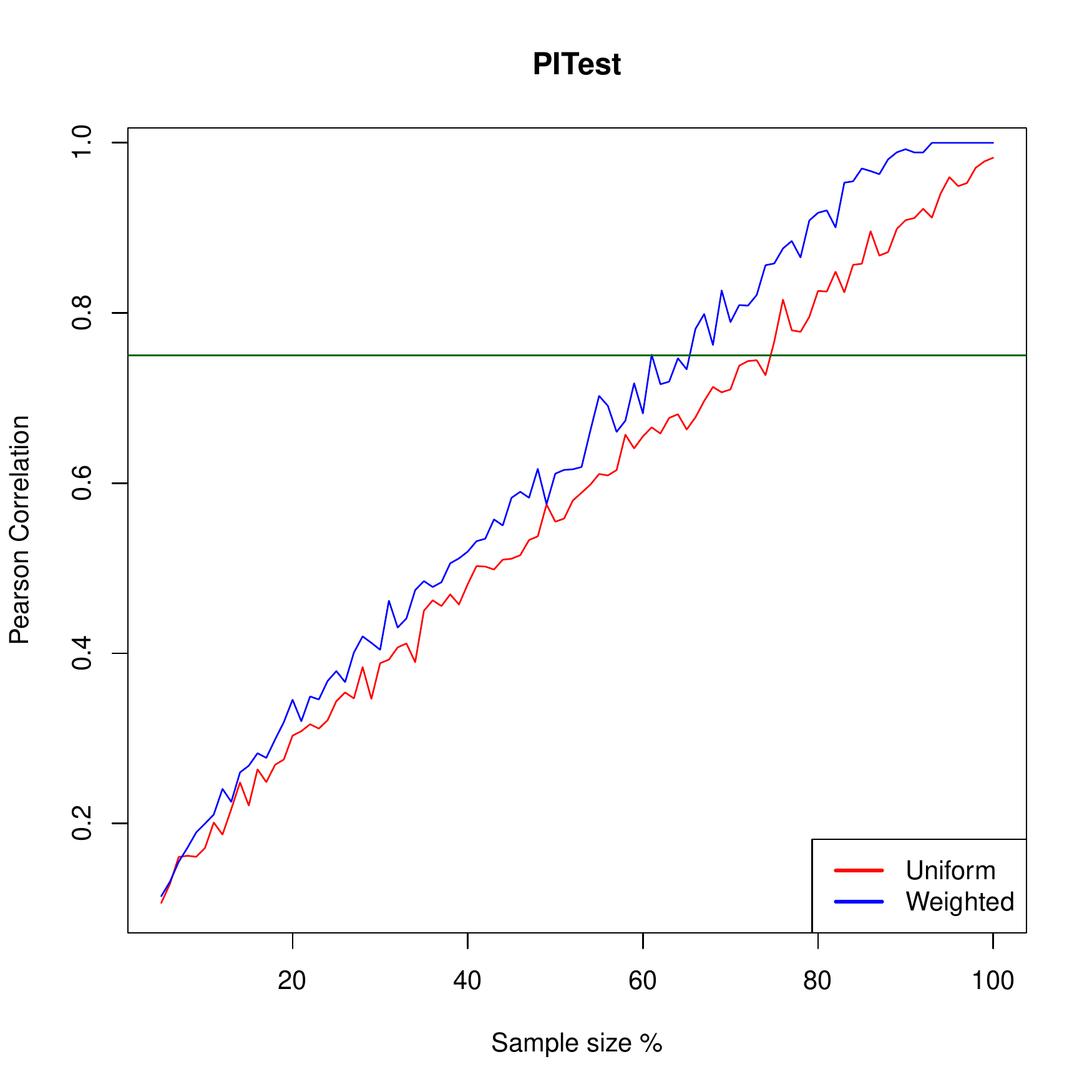}
	\end{minipage}
	\begin{minipage}{0.33\textwidth}
		\centering
		\includegraphics[width=\linewidth]{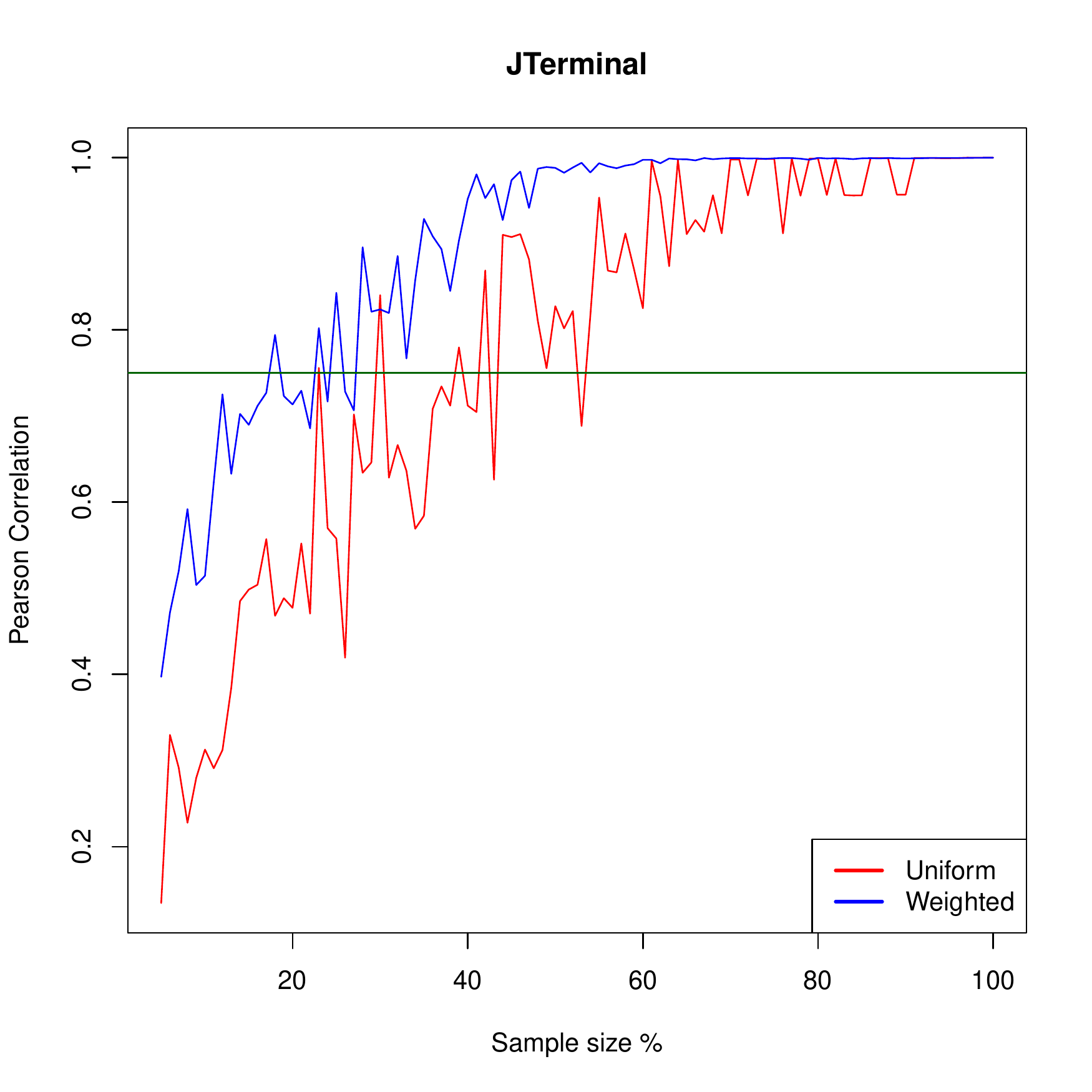}
	\end{minipage}
	\begin{minipage}{0.33\textwidth}
		\centering
		\includegraphics[width=\linewidth]{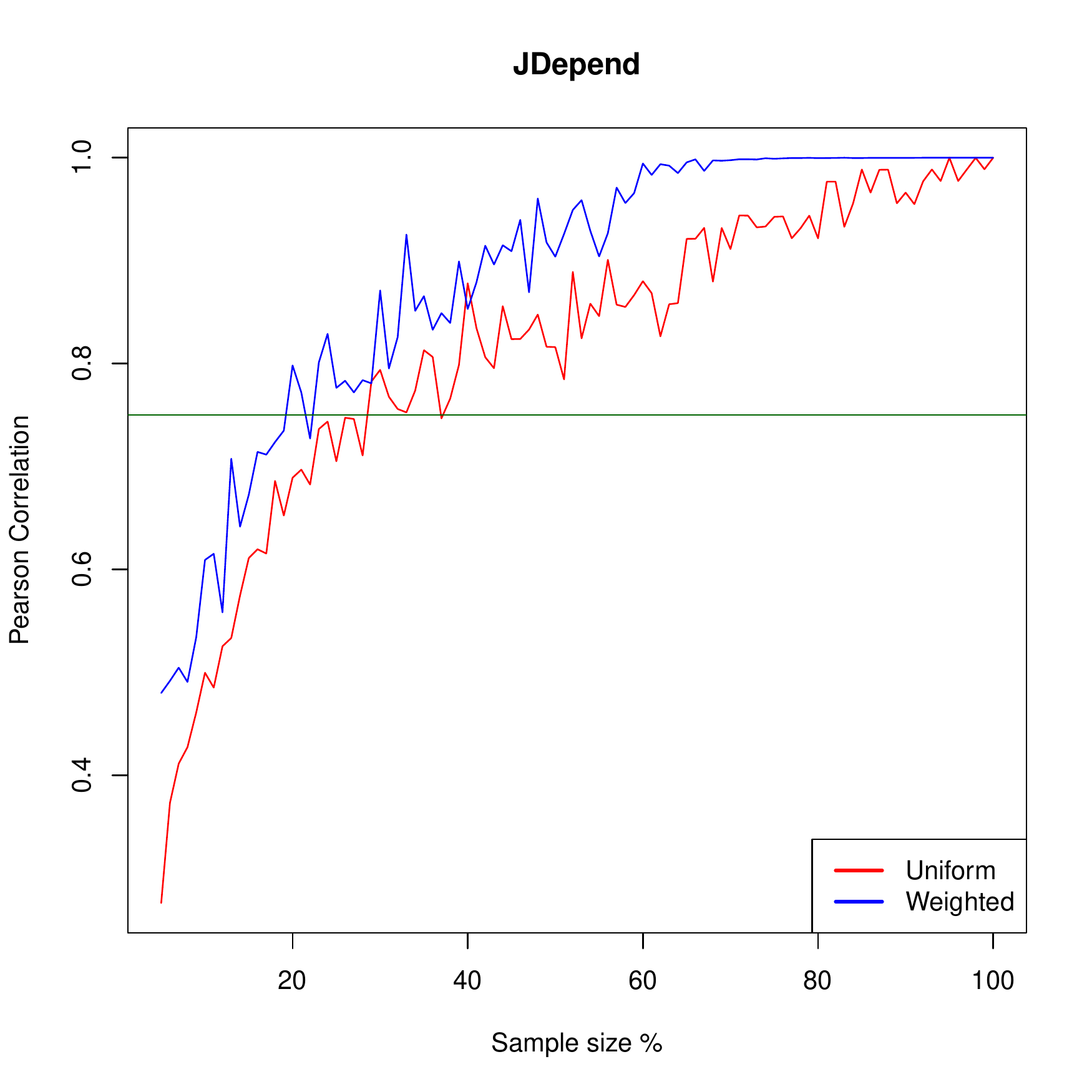}
	\end{minipage}
	\caption{Pearson correlation between sampled and all mutants. The data reported in these figures is discrete. However for better visualization, the points are connected together.}
	\label{fig:pearson}
\end{figure*}

\begin{figure*}
	\begin{minipage}{0.33\textwidth}
		\centering
		\includegraphics[width=\linewidth]{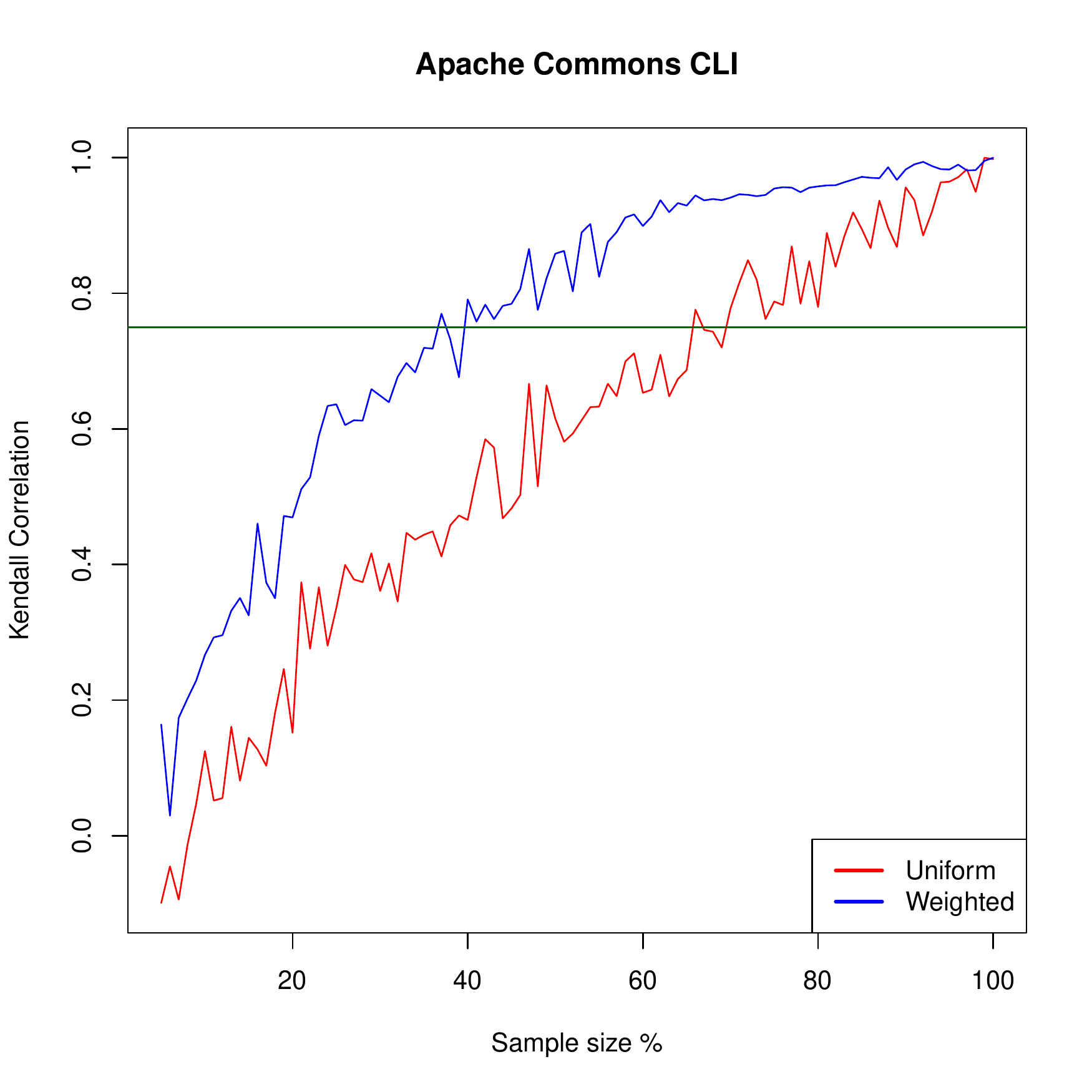}
	\end{minipage}
	\begin{minipage}{0.33\textwidth}
		\centering
		\includegraphics[width=\linewidth]{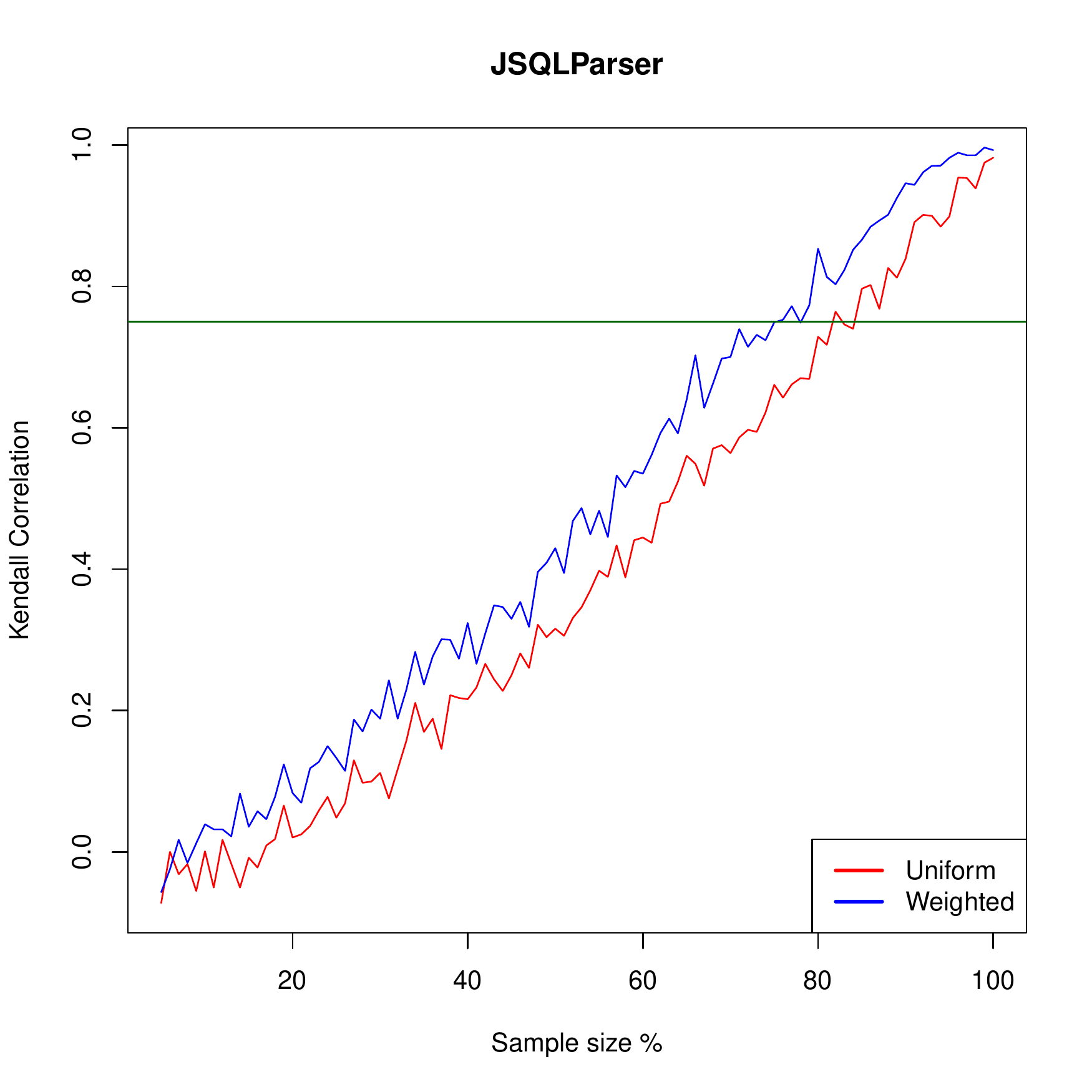}
	\end{minipage}
	\begin{minipage}{0.33\textwidth}
		\centering
		\includegraphics[width=\linewidth]{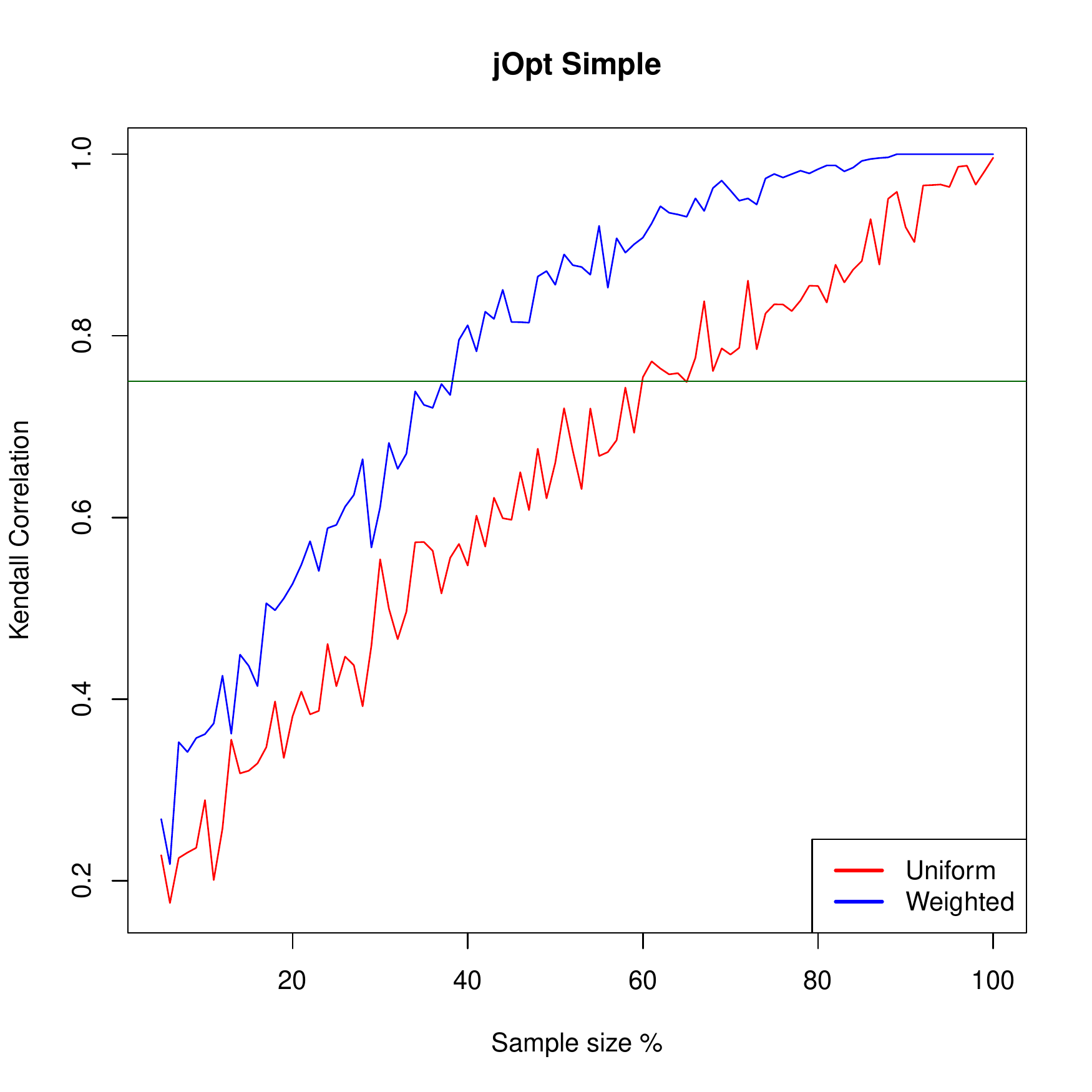}
	\end{minipage}
	\begin{minipage}{0.33\textwidth}
		\centering
		\includegraphics[width=\linewidth]{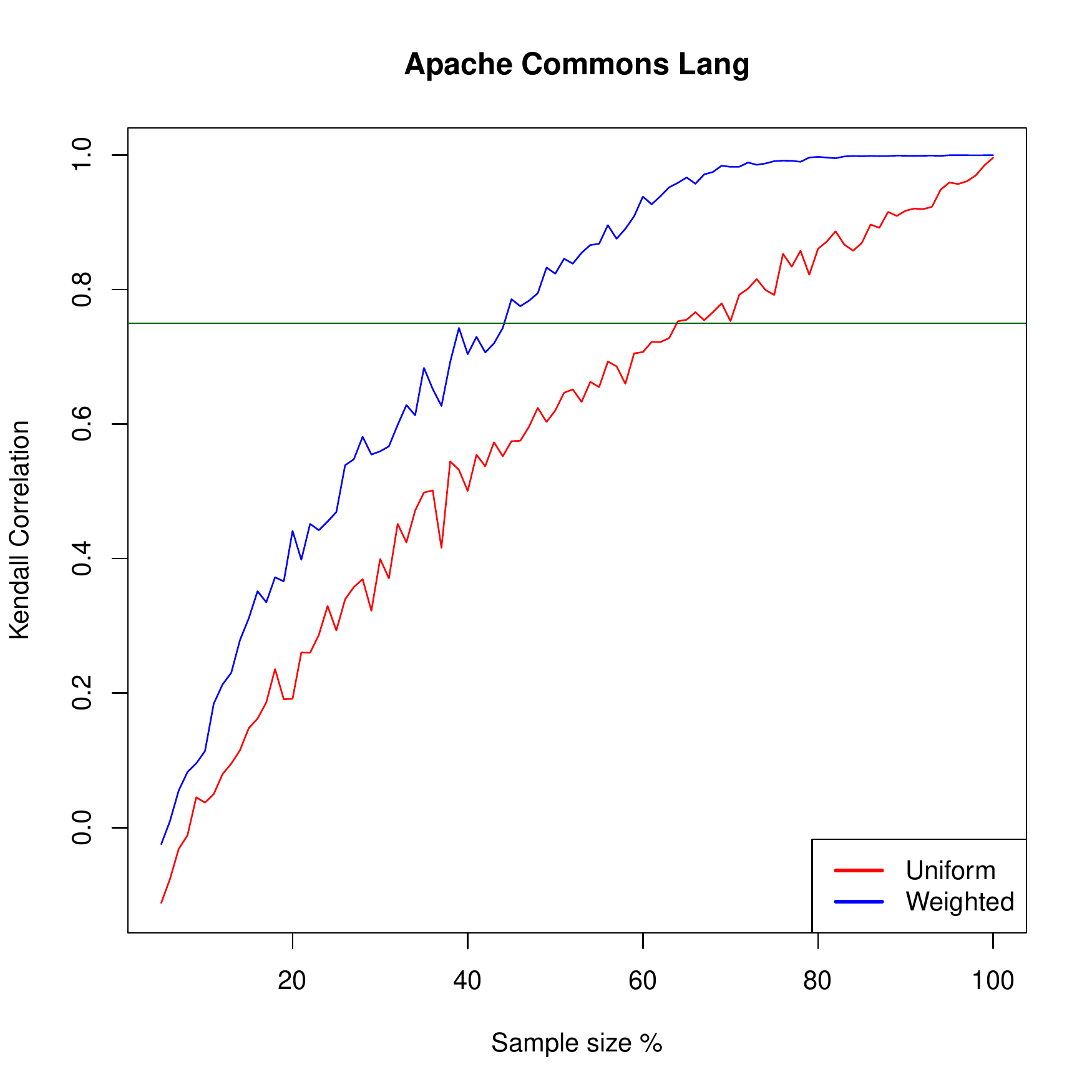}
	\end{minipage}
	\begin{minipage}{0.33\textwidth}
		\centering
		\includegraphics[width=\linewidth]{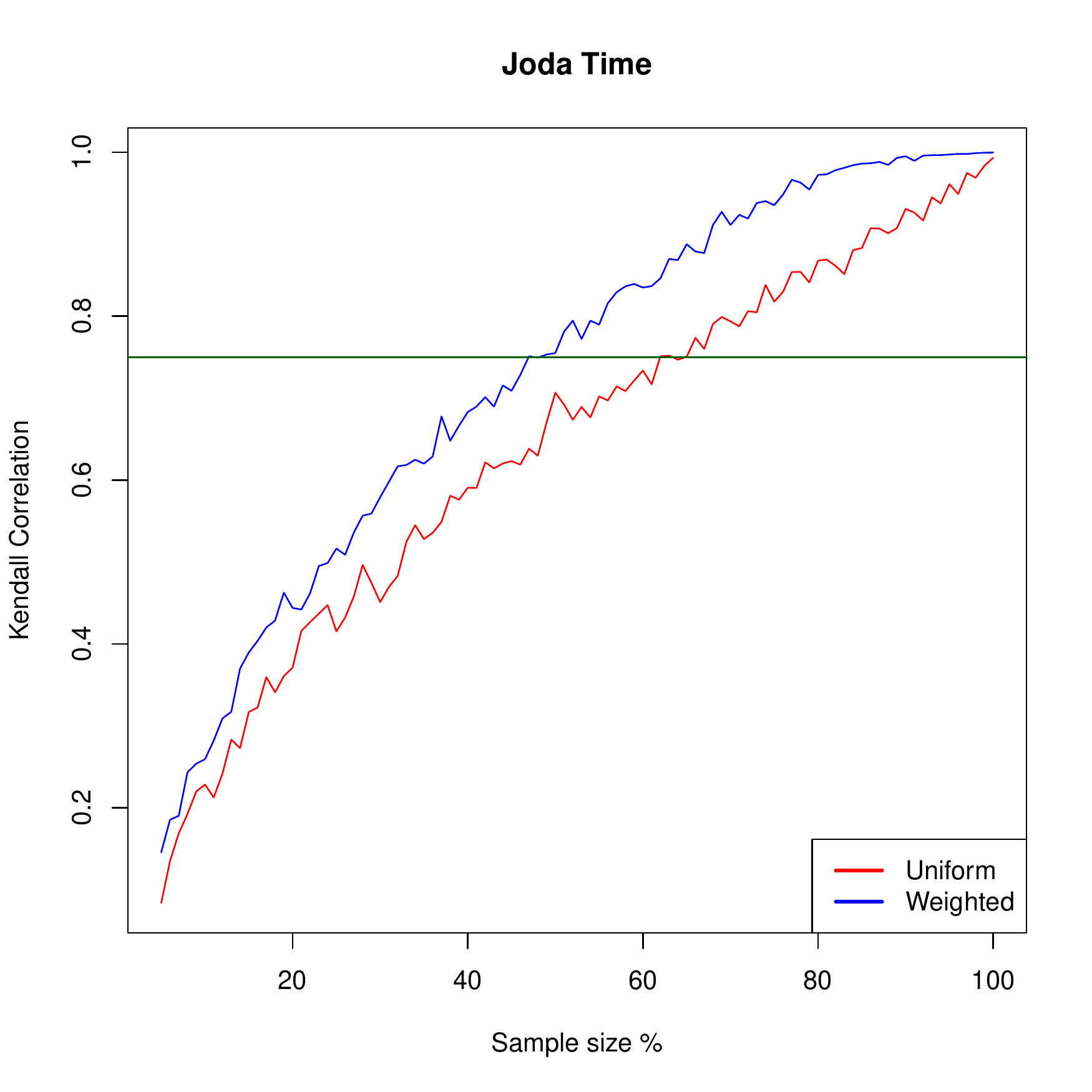}
	\end{minipage}
	\begin{minipage}{0.33\textwidth}
		\centering
		\includegraphics[width=\linewidth]{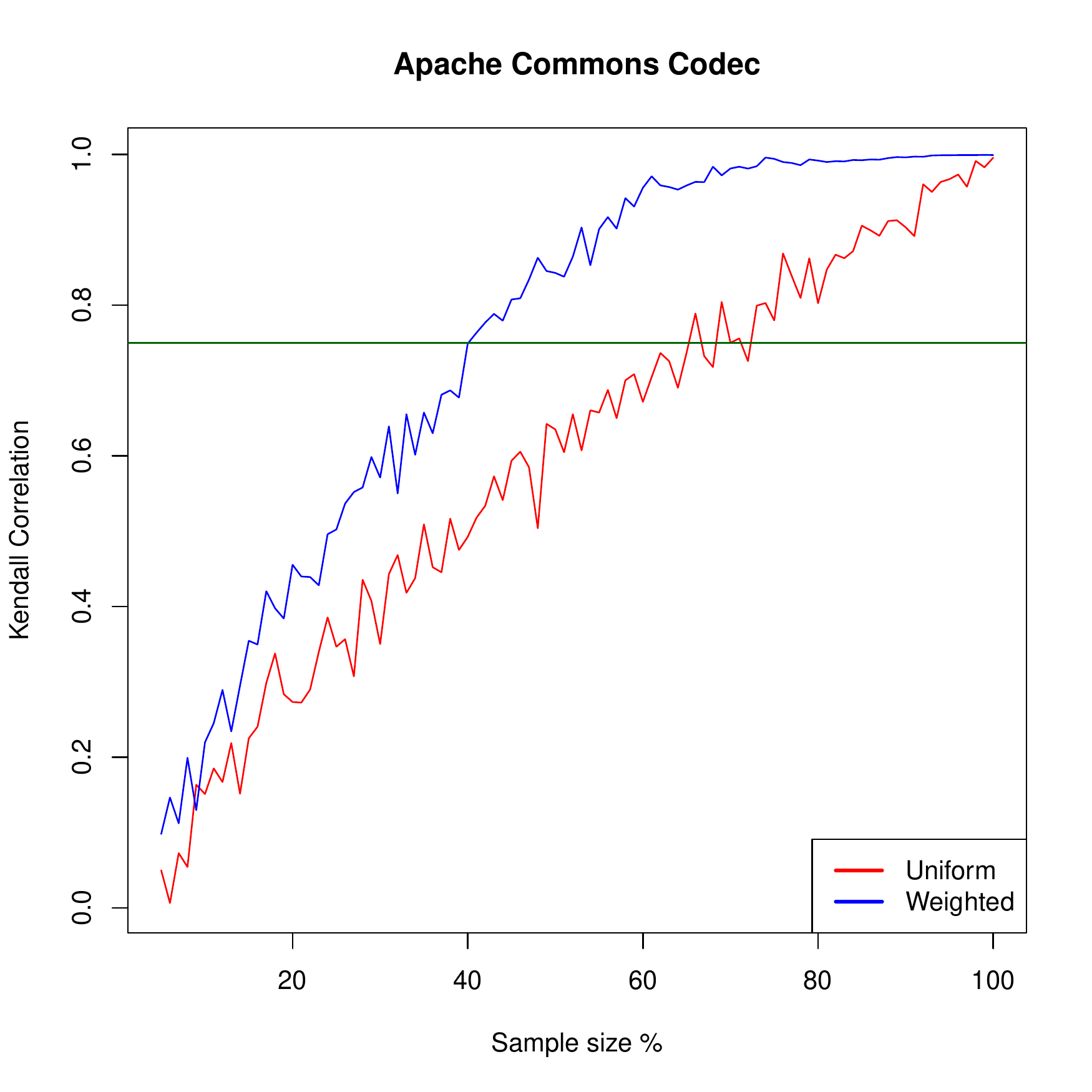}
	\end{minipage}
	\begin{minipage}{0.33\textwidth}
		\centering
		\includegraphics[width=\linewidth]{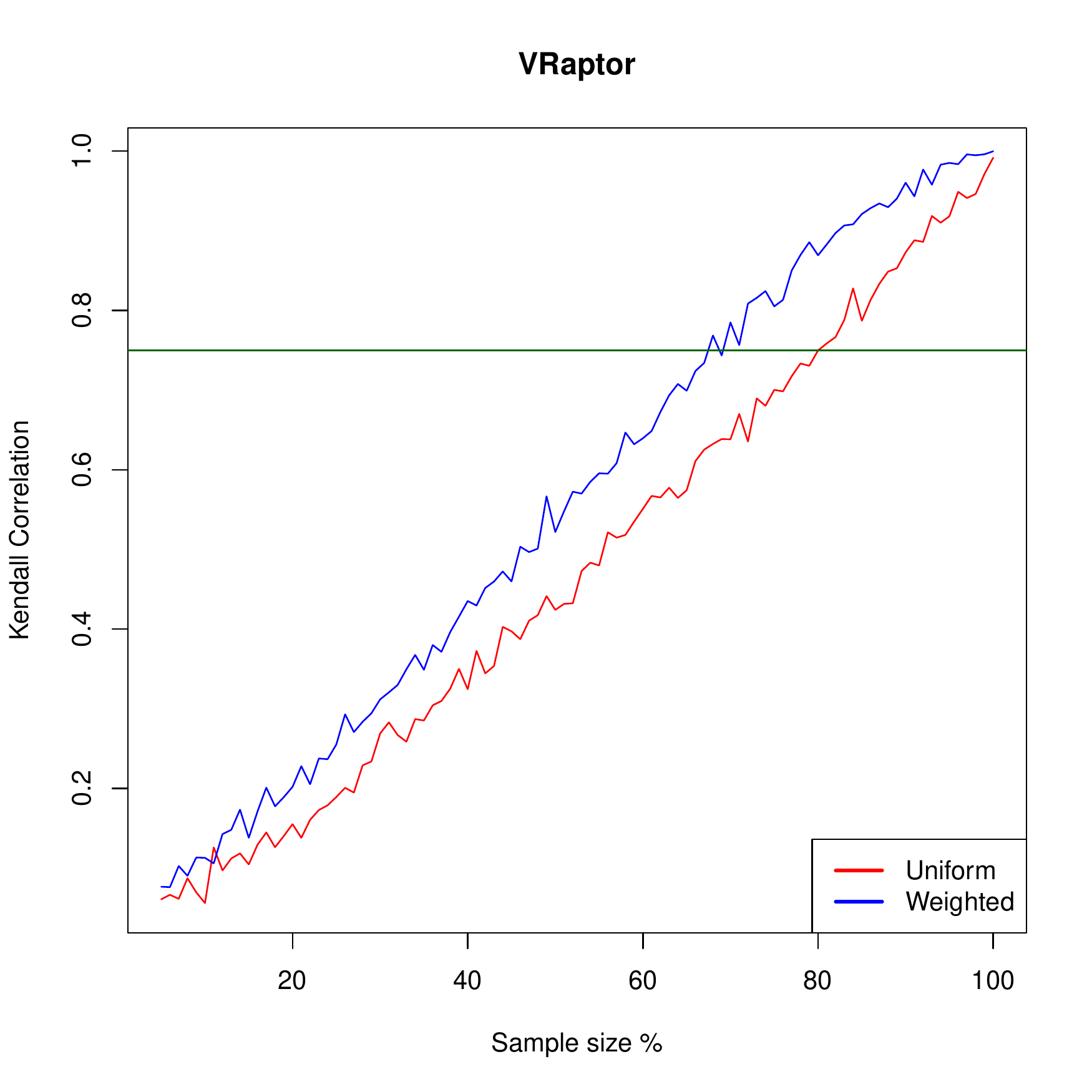}
	\end{minipage}
	\begin{minipage}{0.33\textwidth}
		\centering
		\includegraphics[width=\linewidth]{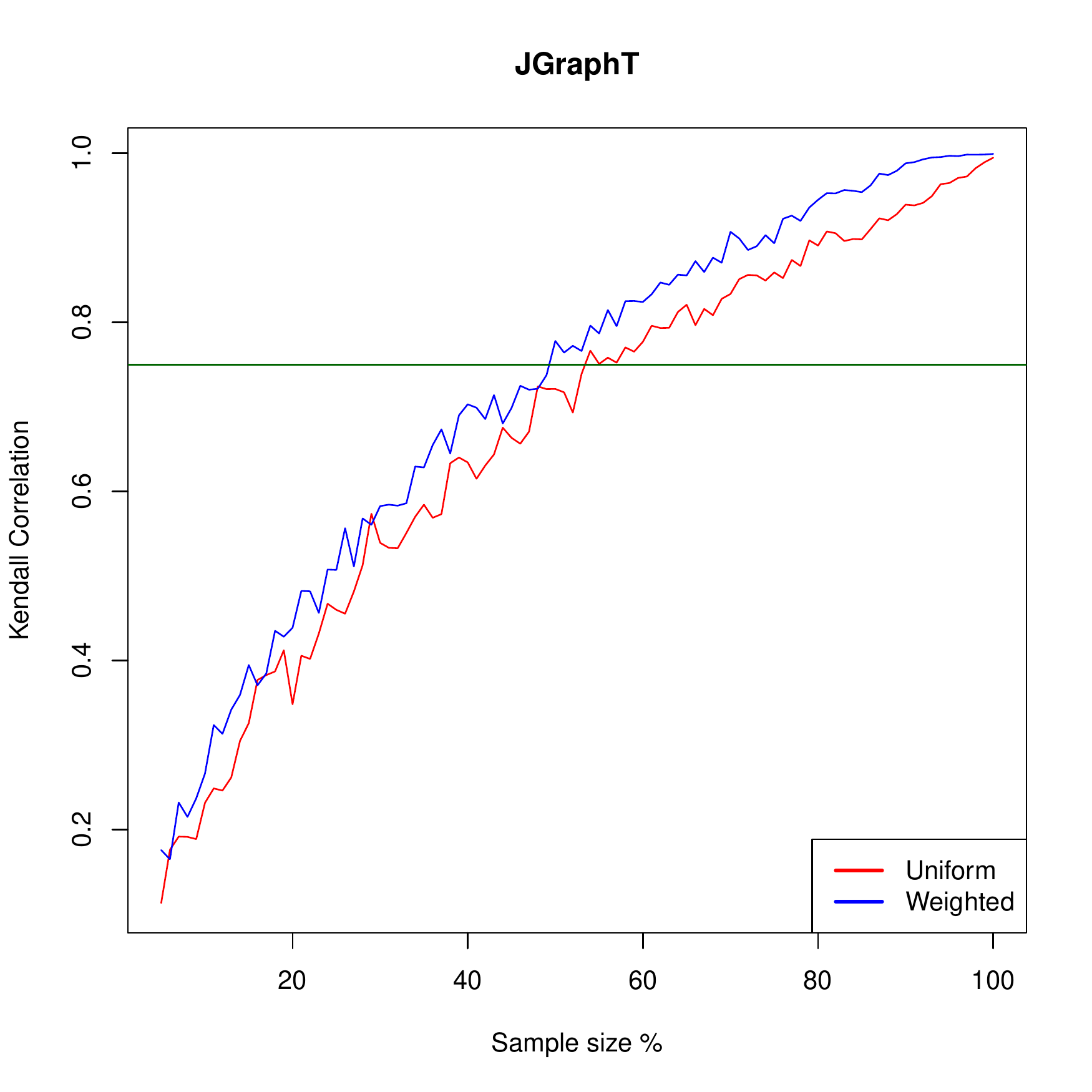}
	\end{minipage}
	\begin{minipage}{0.33\textwidth}
		\centering
		\includegraphics[width=\linewidth]{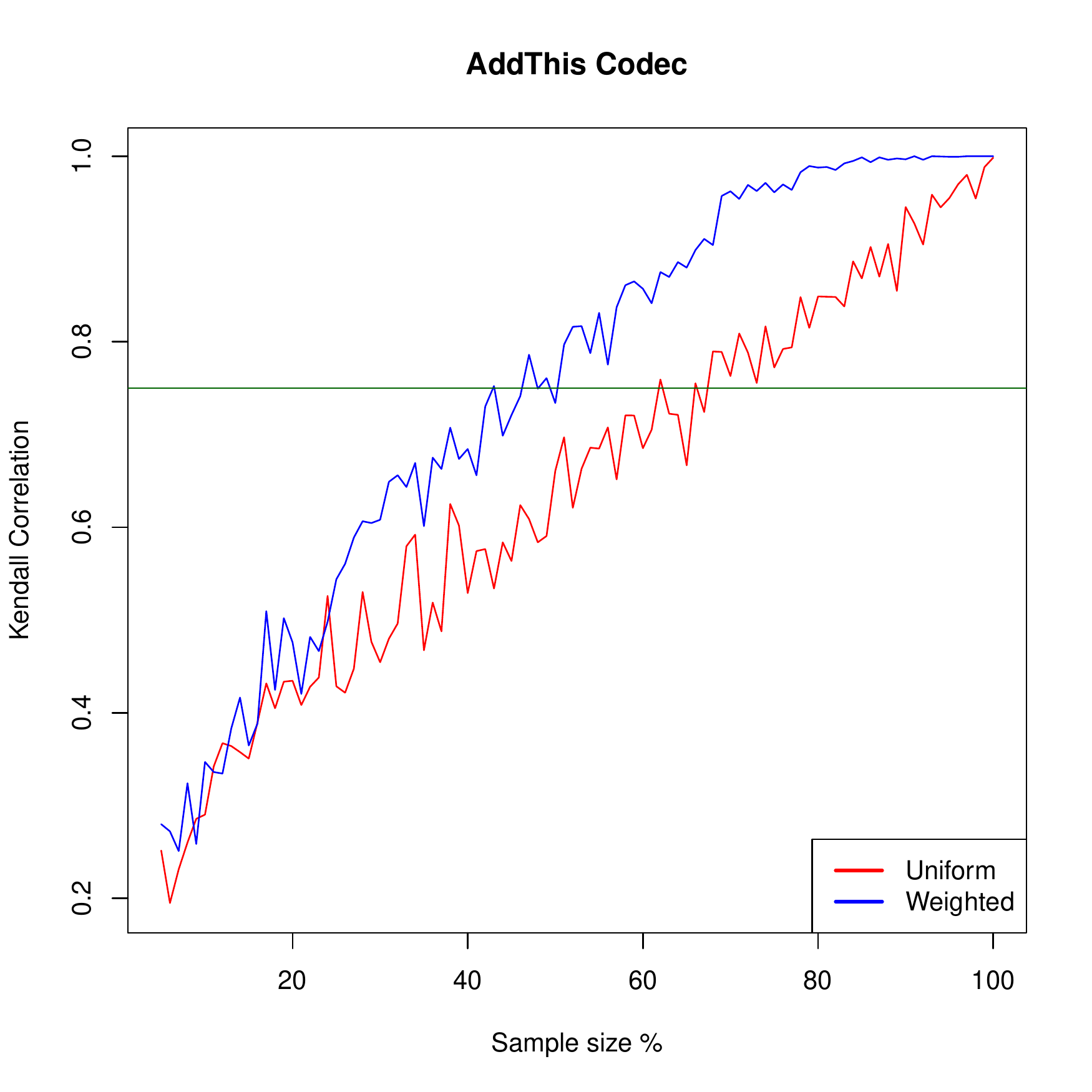}
	\end{minipage}
	\begin{minipage}{0.33\textwidth}
		\centering
		\includegraphics[width=\linewidth]{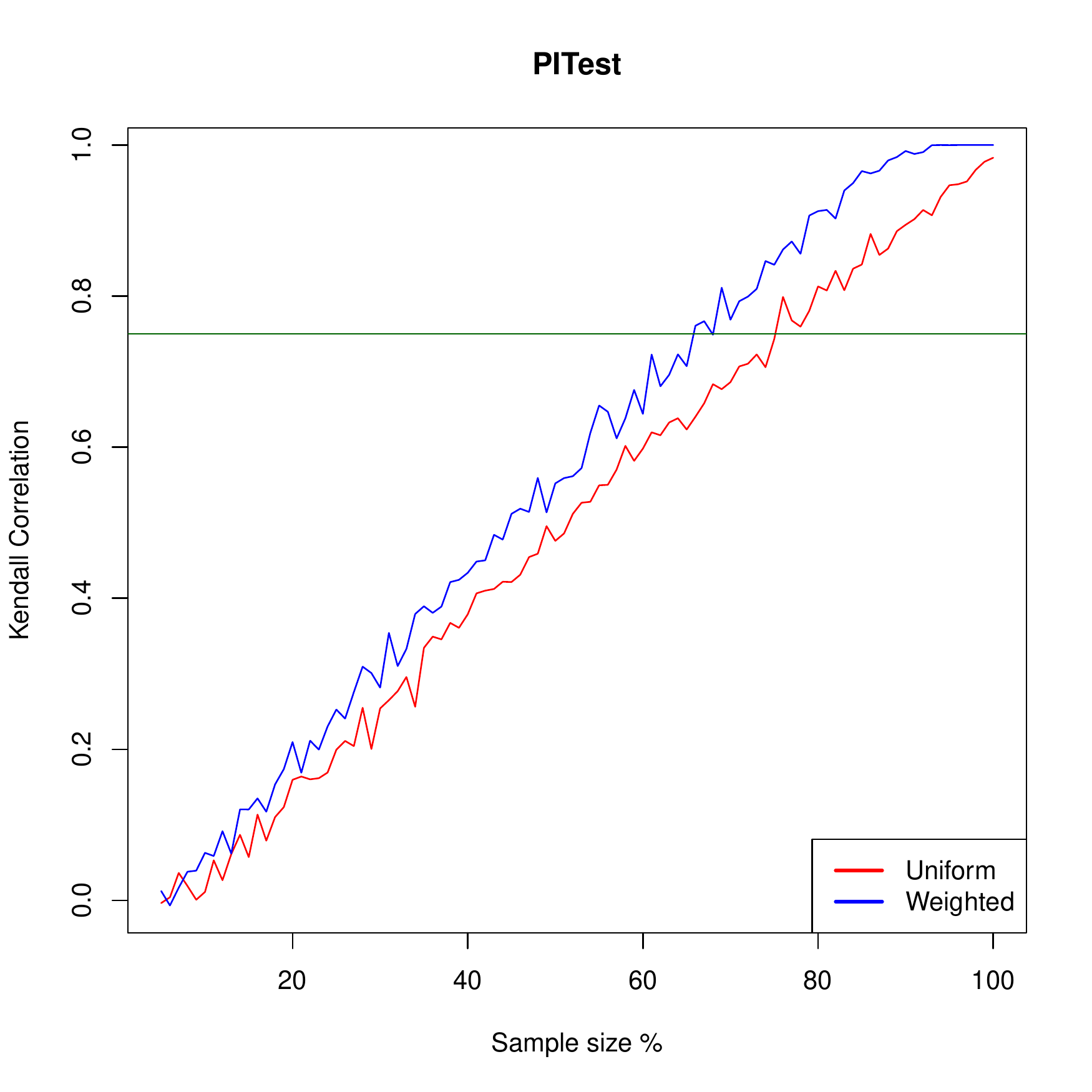}
	\end{minipage}
	\begin{minipage}{0.33\textwidth}
		\centering
		\includegraphics[width=\linewidth]{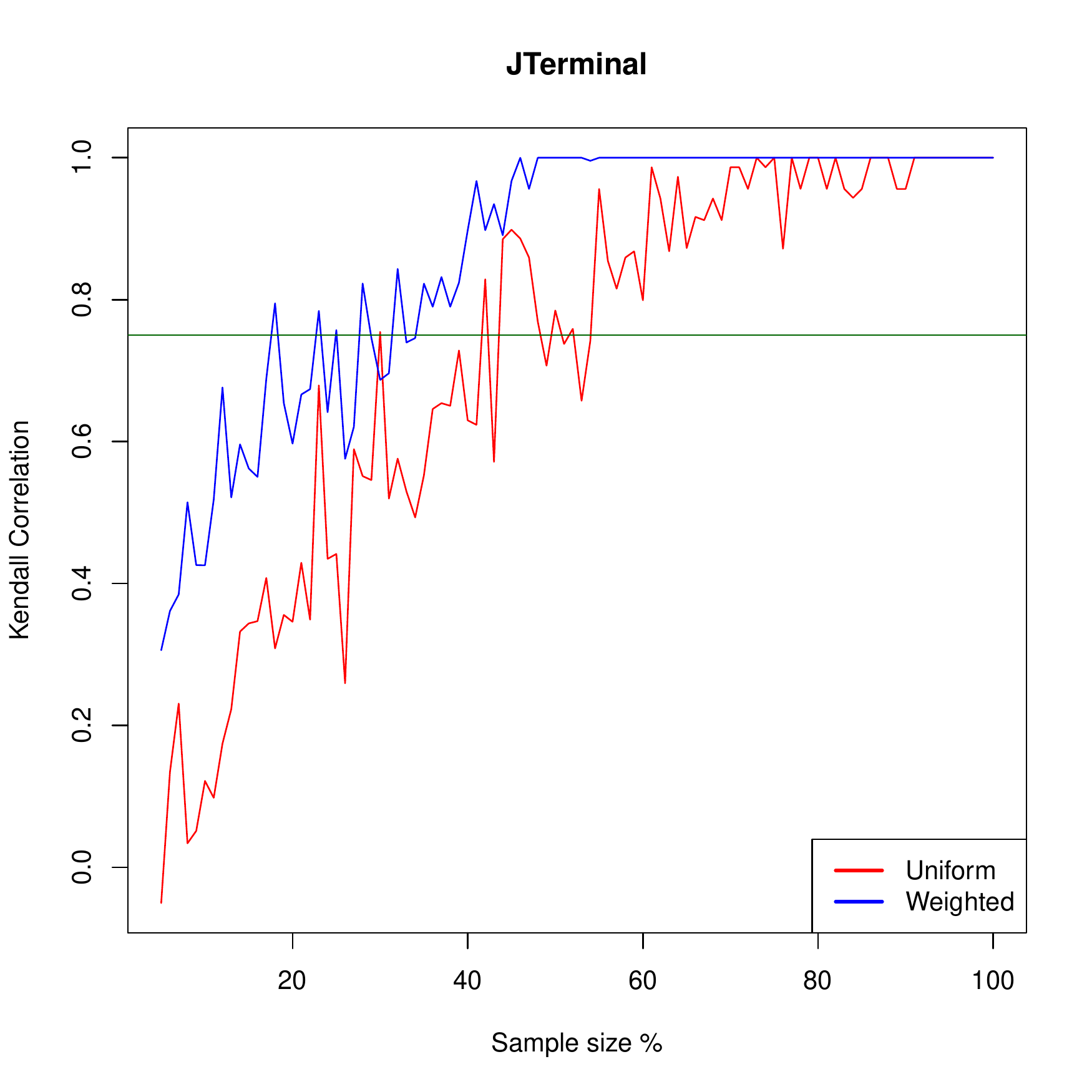}
	\end{minipage}
	\begin{minipage}{0.33\textwidth}
		\centering
		\includegraphics[width=\linewidth]{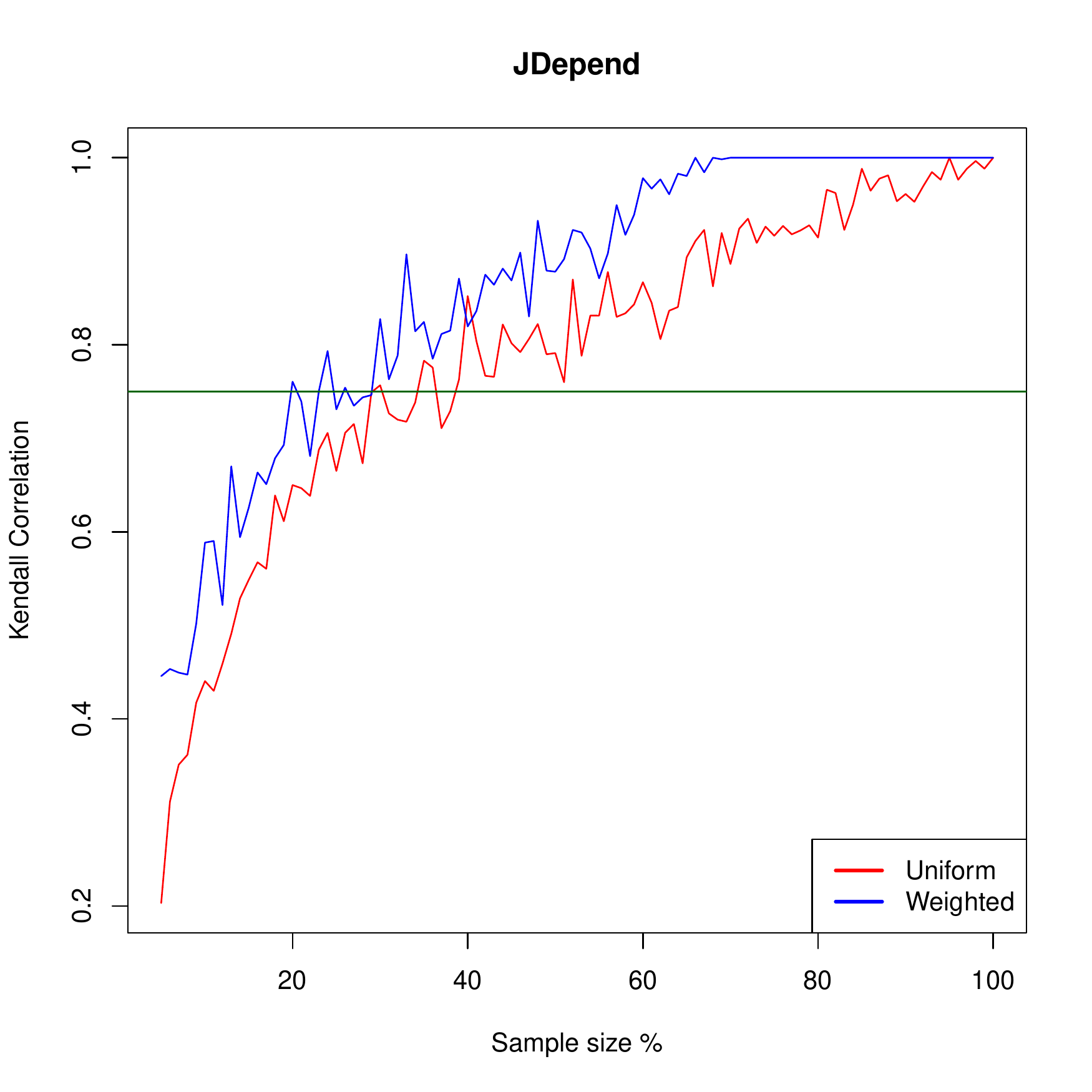}
	\end{minipage}
	\caption{Kendall correlation between sampled and all mutants. The data reported in these figures is discrete. However for better visualization, the points are connected together.}
	\label{fig:kendall}
\end{figure*}

 \textbf{Motivation}. %
  During the analysis of RQ1, we noticed that some classes were not represented at all in the sampled sets of  mutants.   Classes with large number of mutants 
 dominate the sampled set because their mutants have a higher chance of being selected. Consequently, the representativeness  is negatively affected, 
 since the sampled set may not include any mutants from classes with a small number of mutants.
Therefore,  we want a new 
heuristic able to  increase the chances to select mutants from such classes.%
\textbf{Approach}. 
 We introduce a new heuristic that assigns more ``weight'' to the classes with a small number of mutants 
in order to increase the chance to select their mutants. %
Then we analyze to what extent our heuristic reduces the acceptable sampling rate  with respect to the uniform approach (Table~\ref{table:results}).
Finally, we analyze if the acceptable sampling rate using our heuristic is project-dependent. 
For this reason, we analyze the relationship between sampling rate and the degree of representativeness for each project 
(Figures \ref{fig:pearson} and \ref{fig:kendall}).

\textbf{Findings}.  The uniform approach has a degree of representativeness  that grows almost linearly with the sample size (Figures  \ref{fig:pearson} and \ref{fig:kendall}). This means that with a small sample size, it might not produce representative sampled sets. 
On the other hand, for a large sample the reduction in size would be negligible.  
\textbf{Keeping the same degree of representativeness,  the average acceptable sampling rate for the weighted approach is 45\%; which is 18\%  less than the uniform approach}. Almost in every project, the weighted approach goes close to the perfect representativeness around 75\% sampling rate. 
Fixing the sampling rate,  the weighted approach generates sampled sets with higher degree of representativeness compared to the uniform approach (Figures  \ref{fig:pearson} and \ref{fig:kendall}).

\begin{figure}
	\centering
	\begin{minipage}{0.22\textwidth}
\centering
\includegraphics[width=\linewidth]{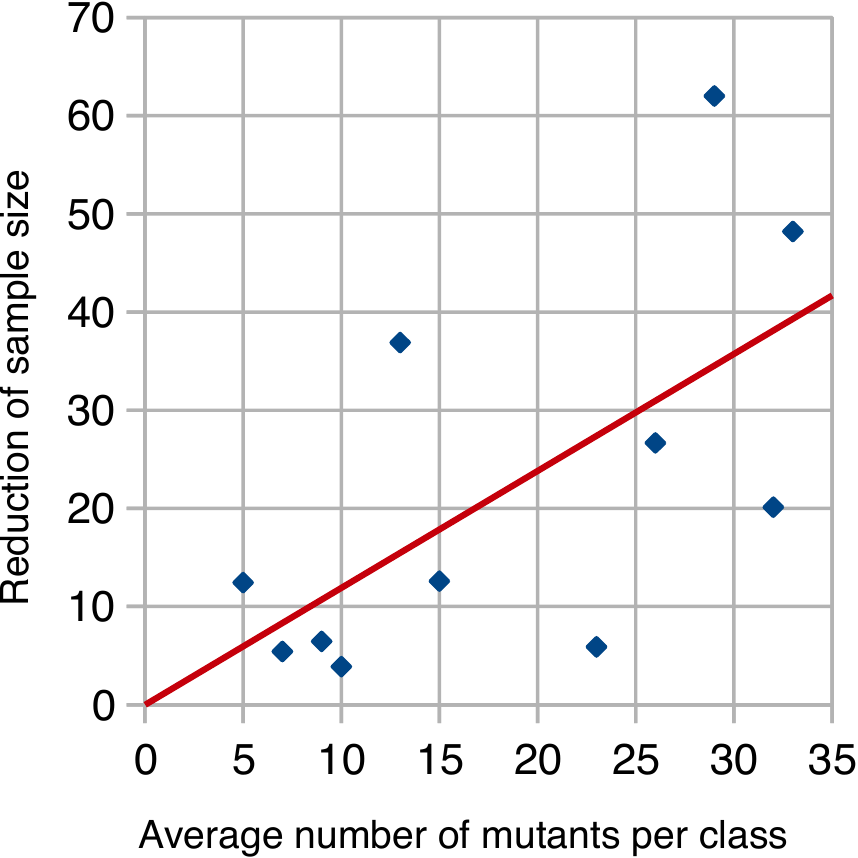}
	\end{minipage}
	\hspace{1mm}
	\begin{minipage}{0.22\textwidth}
	\centering
	\includegraphics[width=\linewidth]{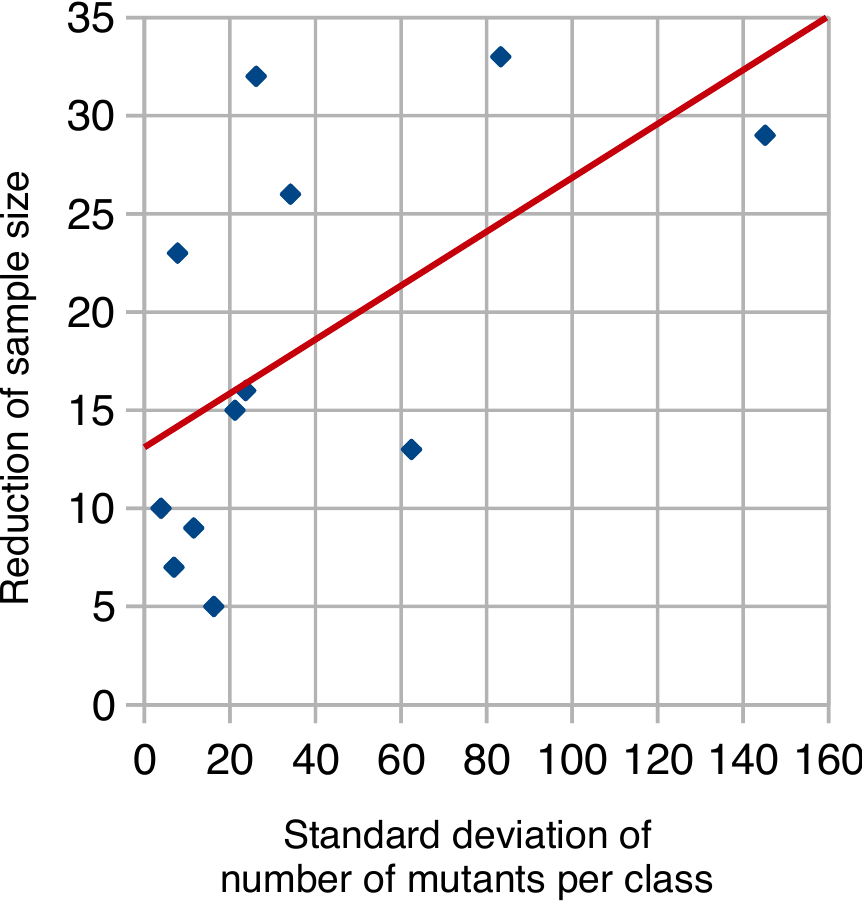}
	\end{minipage}

\caption{How the number of mutants per class affect the reduction of the sample size} %
\label{fig:r3chart}
\end{figure}

\begin{figure}	
		\centering
		\includegraphics[width=0.85\linewidth]{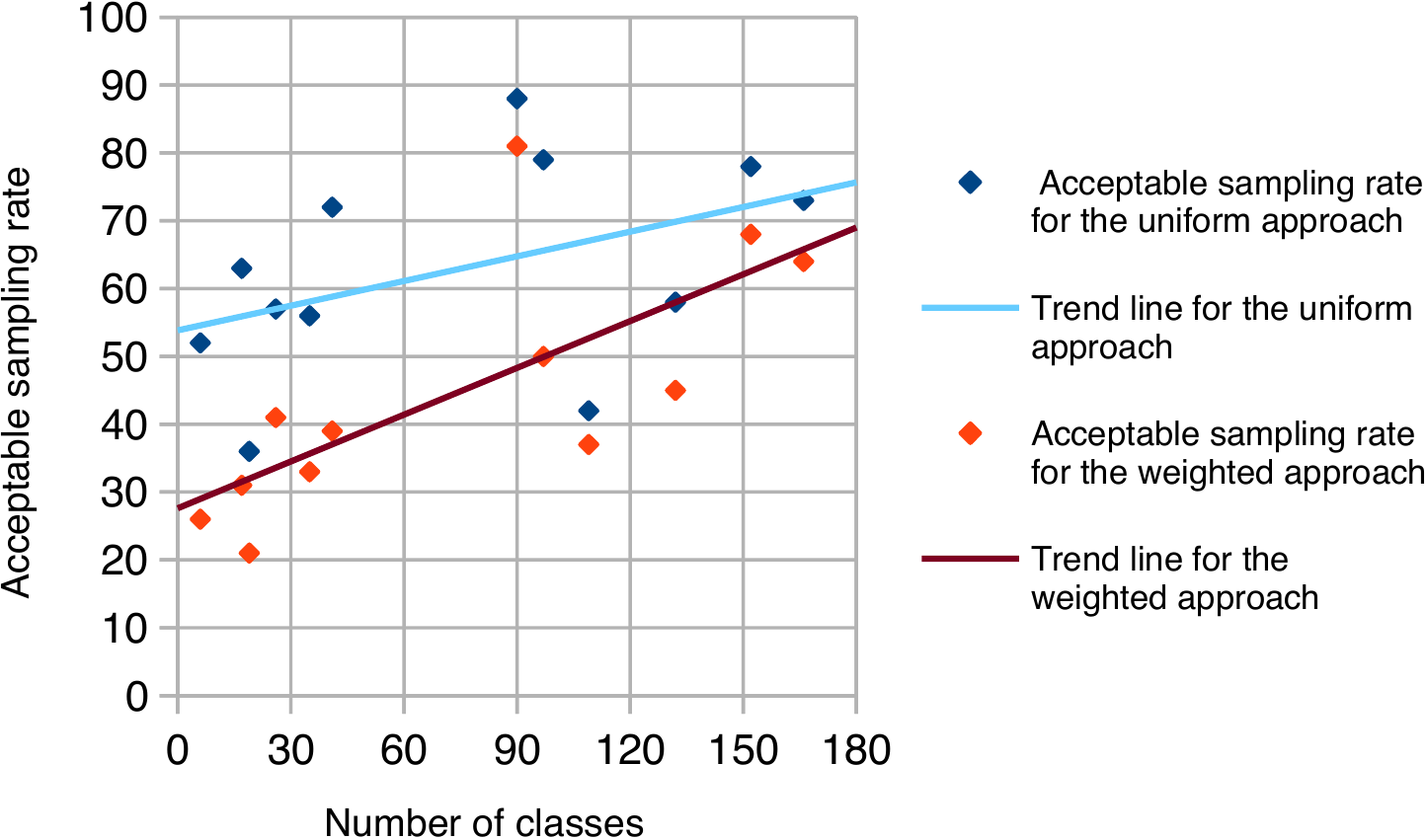}
	\caption{How the number of classes with mutants affects the acceptable sampling rate}%
	\label{fig:RQ3-1}
	
\end{figure}

The reduction in sample size does not follow the same pattern in all projects.
For example, in Apache Commons Codec (second row in the middle in Figures \ref{fig:pearson} and \ref{fig:kendall}), using the uniform approach with a sampling rate higher than 72\%, %
the representativeness  remains acceptable (for both $\rho$ and $\tau_b$).
For the weighted approach, the representativeness  is already acceptable using only a sampling rate of 40\% 
(a reduction of 32\% of  the sample size). 
On the other hand, JGraphT (top left in Figures \ref{fig:pearson} and \ref{fig:kendall}) only shows a 5\% reduction. 
 By investigating this issue further,
 we discovered that for the weighted approach, the correlation between the acceptable sampling rate and the number of classes with mutants is 0.72 with a p-value of 0.008 (Figure~\ref{fig:RQ3-1}), which is higher than the one achieved in the uniform approach. 
 This happens because the more classes there are, the larger the sample needs to be in order to include mutants from all classes. 
 We also find that the  reduction of the sample size has a correlation of 0.57 (p-value 0.053) and 0.64 (p-value 0.025) respectively  with the standard deviation ($\sigma$) and the average of the sizes ($\mu$) of mutant sets for each class (Figure~\ref{fig:r3chart}). 
 This happens because  if the size of classes are close to each other, the weights would be close as well. 
 As a consequence, the weighted approach would behave like the uniform approach.
 Comparing these factors for JGraphT and Apache Commons Codec, we discover that for the former $\sigma$ and $\mu$  are equal to 16.2, and 12.4 respectively, while for the latter  $\sigma$ and $\mu$ are equal to 83.3, and 48.2 respectively. These values are in agreement with our analysis.

\begin{table}
	\centering

		\adjustbox{max width=\linewidth}{	\begin{tabular}{|c|c|c|c|c|}
		\hline \multirow{2}{*}{\textbf{Project}} & \multicolumn{2}{c|}{\textbf{Uniform}} & \multicolumn{2}{c|}{\textbf{Weighted}} \\
		\hhline{~----} & $\rho > 0.75$ & $\tau_b > 0.75$ & $\rho > 0.75$ & $\tau_b > 0.75$ \\ 
		\hline
		\hline Commons CLI	& 63\%	& 69\%	& 31\%	& 39\% \\
		\hline JSQLParser &	88\%	& 83\%	 & 81\%	& 77\% \\ 
 		\hline jOpt Simple &	56\% &	65\%	& 33\% &	38\% \\
		\hline Commons Lang	& 79\%	& 63\%	& 50\%	& 44\% \\
		\hline Joda Time &	58\% &	64\%	& 45\%	& 48\% \\ 
		\hline Commons Codec &	72\% & 	72\% &	39\%	& 40\% \\
		\hline VRaptor &	78\% &	79\% &	68\% &	68\% \\
		\hline JGraphT	& 42\%	& 52\%	& 37\%	& 48\% \\
		\hline AddThis Codec	& 57\%	& 67\%	 & 41\%	& 50\% \\
		\hline PITest	& 73\% &	74\%	& 64\%	& 67\% \\ 
		\hline JTerminal &	52\% &	53\%	& 26\% &	33\% \\
		\hline JDepend	& 36\%	& 37\% &	21\% & 	28\% \\
 				\hline 
		\hline \textbf{Average} &  63\% & 65\%  & 45\%   & 47\% \\
		\hline 
	\end{tabular} }
	\caption{acceptable sampling rate for uniform and weighted approaches}
	\label{table:results}
\end{table}

\begin{figure}[!t]	
	\centering
	\includegraphics[width=0.85\linewidth]{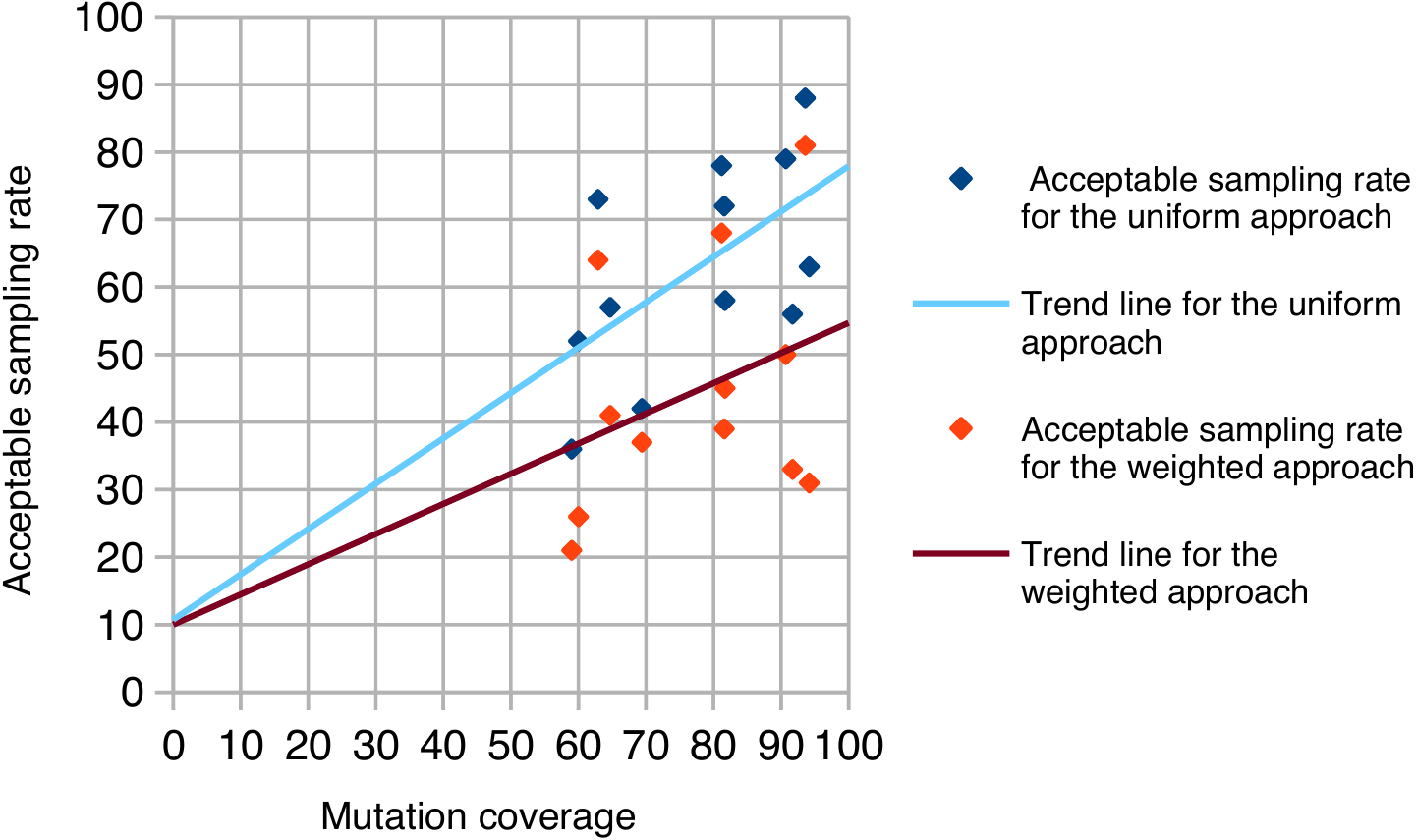}
	\caption{How mutation coverage affect the acceptable sampling rate} %
	\label{fig:RQ3-2}
	
\end{figure}

\subsection*{\textit{RQ3. \RQthree}} 
 \textbf{Motivation}. %
During the investigation of RQ1 and RQ2, we noticed that the acceptable sampling rate is project-dependent. 
In  literature \cite{Wong1993,Wong1995, Zhang2010}, the analysis of sampling rates are done mostly at project level considering projects with adequate test suites.
For this reason, we want to analyze the influence of the level of test adequacy on acceptable sampling rate for each project using class level criteria.

\textbf{Approach}. To evaluate the level of test adequacy we rely on mutation coverage as a metric (Table~\ref{table:cases}). 
Test suites are considered as non-adequate if their mutation coverage is lower than 100\%.
In Table~\ref{table:results},  we sort  projects according to the level of test adequacy to check if and to what extent it influences the acceptable sampling rate.  
In Figure~\ref{fig:RQ3-2}, we plot how test adequacy affects the acceptable sampling rate.

\textbf{Findings}. 
Figure~\ref{fig:RQ3-2} shows that \textbf{for the uniform approach and (to a lesser extent) for the weighted approach, the acceptable sampling rate increases with the test adequacy}.  
This result means that achieving an acceptable degree of representativeness requires higher acceptable sampling rate in projects with higher 
level of test adequacy.
This can be explained considering classes without mutants in the sampled set. 
If these classes had a high level of test adequacy, then their absence in sampled set would have a high negative impact on the degree of representativeness.
 Therefore, a larger sample is needed to be sure that all the classes are represented.  
For example, as seen in Table~\ref{table:results}, JSQLParser and Apache Commons Lang have a acceptable sampling rate of 88\% and 79\% respectively, even though the level of test adequacy is higher than 90\% in both projects.  
This effect is mitigated by using the weighted approach. For this reason, this behavior is less evident than in the uniform approach.

\section{Threats To Validity}
\label{sec:threats}

To describe the threats to validity we refer to the guidelines reported by Yin~\cite{Yin2003}. 

Threats to \textbf{internal validity} focus on confounding factors that can influence the obtained results. 
These threats stem from potential bugs hidden inside the algorithms used for sampling mutants or in LittleDarwin. 
We consider this chance ---even if possible--- limited. The pseudo-code for random  mutant selection is explained in our paper and its implementation has been carefully reviewed by the first author.  %
The code of LittleDarwin has been already checked and tested in several case studies~\cite{Parsai2015,Parsai2015a}. %
 Finally, the code of LittleDarwin along with all the raw data of the study is publicly available for download in the replication  package.\footnote{\url{http://parsai.net/files/research/ReplicationPackage.7z}} %

Threats to \textbf{external validity} correspond to the generalizability of our results. In our analysis we use only 12 open source projects. We mitigate this threat
by using projects  which differ for number of contributors, size and adequacy of the test suite. Yet,  it is desirable to replicate this study using more projects, 
especially the ones belonging to industrial settings. A second threat stems from the limited set of mutation operators used in mutation testing process. 
Since mutation coverage depends on the mutants used, the sampling process and its accuracy is bounded to mutation operators that generate them. The impact of this threat is limited since 
we use the standard set of mutation operators which is representative of mistakes commonly introduced by developers and typically supported by many mutation testing tools~\cite{Parsai2015}. In addition, since both uniform and weighted random selection was performed using the same set of mutation operators, the impact of extra  mutation operators on the results of the study is minimal. 

Threats to \textbf{construct validity}  are concerned with how accurately the observations describe the phenomena of interest. In our case, this depends on the set of metrics adopted to evaluate the algorithms for random mutant  selection.
To measure to what extent the sampled mutants are representative of all possible mutants, we use the correlation between mutation coverage values calculated using sampled set and full set of mutants, thus emulating the way coverage metrics are commonly used in industry. We calculate correlations using well known metrics that have been used in literature numerous times. 
Even though we define the critical point after which the correlation is strong  at 0.75, our results still hold if we choose slightly different thresholds. %

Sampling at project-level is another threat to construct validity. However, the alternative of sampling at class-level was not viable, since it is not possible to guaranty a certain size for the sampled set at project level. 

As for filtering the equivalent mutants, we chose not to categorically remove a group of mutants, because the distribution of equivalent mutants in the sampled sets theoretically remains the same as the distribution in all mutants. Since, equivalent mutants act as false positives~\cite{Grun2009},  we can argue that our sampled sets contain roughly the same percentage of false positives as in all mutants. Because classifying  a single mutant for a class that has only few mutants  as a false negative is more costly than adopting few false positives, we prefer the latter based on common practice~\cite{Fawcett2006}. 

Threats to \textbf{conclusion validity} are concerned with the degree to which conclusions we reach about relationships in our data are reasonable. Since the provided rationals in Section~\ref{sec:ar} justify our conclusions, we assume there are no threats to conclusion validity.

\section{Related Work}
\label{sec:rw}

Reduction of the number of mutants has been investigated in literature to reduce the computational cost of mutation testing. Mutant selection is a simple approach towards this goal.
There are two main branches to mutant selection: %
operator-based mutant selection and random mutant selection.

Operator-based mutant selection has been studied in detail in literature. %
Mathur~\cite{Mathur1991} first proposed an approach based on selecting the \textit{sufficient} set of mutant operators. Wong et al.~\cite{Wong1995} examined a selective set of mutation operators (2 out of 22) and concluded that the results are similar to those of all mutation operators. Offutt et al.~\cite{Offutt1993,Offutt1996} demonstrated through empirical experiments that  five mutation operators are sufficient to emulate the full set of mutation operators. Barbosa et al. uses random mutant selection as a control technique to determine the sufficient set of mutation operators for C~\cite{Barbosa2001}. Siami Namin et al.~\cite{SiamiNamin2008} used a subset of mutation operators for Proteum~\cite{Maldonado2001} that generated only 8\% of all mutants. Gligoric et al.~\cite{Gligoric2013} extended this topic to concurrent code, and propose a set of 6 mutation operators as the sufficient set for the concurrent code.  %

Random mutant selection has not been investigated as deeply as operator-based mutant selection in the literature. The first mention of random mutant selection is in the works of Acree et al.~\cite{Acree1979,Acree1980} and Budd~\cite{Budd1980}, however, they did not perform any extensive empirical research on this subject.  %
Wong and Mathur~\cite{Wong1995} then examined the uniform approach with incremental steps and concluded that a selection rate of 10\% is only 16\% less effective than all mutants. Zhang et al. have shown that the uniform approach using half the mutants is as effective as using all mutants~\cite{Zhang2010}.  They also %
proposed a two-round random mutant selection in which first a mutation operator is selected, and then a mutant generated by that operator is randomly selected. They demonstrated that this approach is more reliable due to less variance in the selected random sets in different runs. %
However, they used only projects with adequate test suites as the subjects of their experiment. 
Zhang et al. show that the use of random mutant selection along with operator-based mutant selection produces more accurate results~\cite{Zhang2013}. They also show that the  uniform approach provides accurate overall mutation coverage values for the test suite. In these studies~\cite{Wong1995,Zhang2010,Zhang2013} they only consider the correlation between several sets of test suites calculating the overall mutation coverage for the project. %
Our study differs from the previous ones in these respects:

\begin{compactitem}
	\item Instead of analyzing the representativeness of the sampled set at project level, we analyze it at class level. 
	In this way, we avoid 
	domination of larger classes over smaller ones in the analysis of the representativeness at project level. 
	\item We study the largest set of projects with non-adequate test suites. Such projects have different characteristics such as size, number of contributors, and level of test adequacy. 
	
\end{compactitem}

\section{Conclusion and Future Work}
\label{sec:conclusions}

Mutation testing is a widely studied method to determine the adequacy of a test suite. However, its adoption in real scenarios is hindered by its computationally intensive nature. 
Several approaches have been proposed to make it feasible in industrial settings and among them random mutant selection shows promising results.
Studies of random mutant selection have two shortcomings, 
they focus their analysis at project level and they are mainly based on projects with adequate test suites. 
In this study we attempt to fill this the lack of empirical evaluation.
We analyze uniform and weighted approaches in the context of random mutant selection.
We compare these approaches at class level using as baseline 12 projects with 
various levels of test suite adequacy, code base sizes, and contributors.

We highlight that the uniform approach underachieves the expected results when analyzed at class level. 
We show that the uniform approach is only viable using a sampling  rate around 65\% (on average). Moreover,
the degree of  representativeness of the sampled sets grows linearly with the increase of the sampling rate. 
We also show that the average acceptable sampling rate for the weighted approach  is 18\% less than the uniform approach while keeping the same degree of representativeness.
The reduction in sample size between these two approaches is correlated with the average, 
and the standard deviation of the number of mutants per class.  
We discover that acceptable sampling rate is correlated with the number of classes with mutants in the uniform (and to a higher extent) to the weighted approach. 
We discover that the lack of representation of a class with a small set of mutants in the sampled set affects the projects with higher test adequacy in a stronger manner. 
By using the weighted approach this  problem is reduced since it increases the chance of inclusion of mutants from classes with small set of mutants in the sampled set. %
	
We discovered that the number of mutants per class is a relevant factor to create a representative sampled set. For this reason, we invite fellow researchers to explore the influence of other factors such as type and position of the mutants for increasing the degree of representativeness.

\section*{Acknowledgments}
This work is sponsored by the Institute for the Promotion of Innovation through Science and Technology in Flanders through a project entitled Change-centric Quality Assurance (CHAQ) with number 120028.

\balance

\printbibliography

\newpage

\balance

\end{document}